\documentclass{article}
\usepackage{graphicx}
\usepackage{amsmath}
\usepackage{amsfonts}
\usepackage{amssymb}

\begin{document}

\author{Antony Valentini\\Augustus College}

\begin{center}
{\LARGE Primordial quantum nonequilibrium and large-scale cosmic anomalies}

\bigskip

\bigskip

\bigskip

\bigskip

Samuel Colin and Antony Valentini\footnote{Corresponding author
(antonyv@clemson.edu)}

\textit{Department of Physics and Astronomy,}

\textit{Clemson University, Kinard Laboratory,}

\textit{Clemson, SC 29634-0978, USA.}

\bigskip

\bigskip
\end{center}

\bigskip

\bigskip

\bigskip

\bigskip

\bigskip

\bigskip

We study incomplete relaxation to quantum equilibrium at long wavelengths,
during a pre-inflationary phase, as a possible explanation for the reported
large-scale anomalies in the cosmic microwave background (CMB). Our scenario
makes use of the de Broglie-Bohm pilot-wave formulation of quantum theory, in
which the Born probability rule has a dynamical origin. The large-scale power
deficit could arise from incomplete relaxation for the amplitudes of the
primordial perturbations. We show, by numerical simulations for a spectator
scalar field, that if the pre-inflationary era is radiation dominated then the
deficit in the emerging power spectrum will have a characteristic shape (an
inverse-tangent dependence on wavenumber $k$, with oscillations). It is found
that our scenario is able to produce a power deficit in the observed region
and of the observed (approximate) magnitude for an appropriate choice of
cosmological parameters. We also discuss the large-scale anisotropy, which
might arise from incomplete relaxation for the phases of the primordial
perturbations. We present numerical simulations for phase relaxation, and we
show how to define characteristic scales for amplitude and phase
nonequilibrium. The extent to which the data might support our scenario is
left as a question for future work. Our results suggest that we have a
potentially viable model that might explain two apparently independent cosmic
anomalies by means of a single mechanism.

\bigskip

\bigskip

\bigskip

\bigskip

\bigskip

\bigskip

\bigskip

\bigskip

\bigskip

\bigskip

\bigskip

\bigskip

\section{Introduction}

According to inflationary cosmology, the observed anisotropies in the cosmic
microwave background (CMB) were ultimately seeded by primordial quantum
fluctuations \cite{LL00,Muk05,W08,PU09}. Precision measurements of the CMB may
therefore be interpreted as tests of quantum mechanics -- as well as of
fundamental physics generally -- at very early times and at very short
distances \cite{PSS06,AV07,AV08,AV10,MVP12,LSS12,CPS13,DLSS13,CV13}. In this
paper we focus on a scenario in which the quantum Born probability rule may
have been violated at very early times, resulting in corrections to the
primordial spectrum at very large wavelengths \cite{AV07,AV08,AV10,CV13}. This
scenario is natural in the de Broglie-Bohm pilot-wave formulation of quantum
theory \cite{deB28,BV09,B52a,B52b,Holl93}, in which it has been argued that
the Born rule is not a law but only a particular state of statistical
equilibrium
\cite{AV91a,AV91b,AV92,AV96,AV01,AV02,AV07,AV08,AV09,AV10,AVPwtMw,PV06}. In a
cosmology with a radiation-dominated pre-inflationary phase
\cite{VF82,L82,S82,PK07,WN08}, if the universe is assumed to begin in a state
of `quantum nonequilibrium' with a statistical spread smaller than that
implied by the Born rule, then on expanding space the dynamics yields
efficient relaxation to equilibrium at short (sub-Hubble) wavelengths and a
suppression or retardation of relaxation at long (super-Hubble) wavelengths
\cite{AV07,AV08,AV10,CV13}. It is then a natural prediction of pilot-wave
theory that, at the onset of inflation, the primordial spectrum will show an
anomalous deficit at sufficiently long wavelengths \cite{AV07,AV08,AV10,CV13}.

Data from the \textit{Planck} satellite appear to show a power deficit of
5--10\% in the multipole region $l\lesssim40$, with a statistical significance
in the range 2.5--3$\sigma$ (depending on the estimator) \cite{PlanckXV}. The
statistical significance is not high, but nevertheless (as the Planck team has
noted) it is important to consider theoretical models that predict a low-$l$
deficit, in order to better assess the potential significance of this finding.
A related anomaly concerns the (temperature) two-point angular correlation
function at large scales, which is smaller than expected with a statistical
significance exceeding 3$\sigma$ \cite{CHSS13}.

It is conceivable that the observed deficit is caused by an incomplete
relaxation to quantum equilibrium during a pre-inflationary era (though of
course it might be caused by some other effect). The measured wavelength at
which a relaxation-induced power deficit could set in will depend on unknown
cosmological parameters, in particular the number of inflationary e-folds. It
is possible that the purported effect exists, but at wavelengths too large to
be observable. On the other hand, should the effect exist in an observable
range, what particular signatures would it display? That is the subject of
this paper. We perform extensive numerical simulations of quantum relaxation
for a spectator scalar field on a radiation-dominated (purportedly
pre-inflationary) background, for varying wavelengths, as well as for varying
numbers of excited states and for varying time intervals. Our aim is to find
features of the corrected spectrum that are broadly independent of the precise
(and unknown) details of the putative pre-inflationary era -- features that,
in future work, could be subjected to a rigorous statistical comparison with
data. We find in particular that the primordial spectrum will be diminished by
a factor $\xi(k)$ that is predicted to be an inverse-tangent function of
wavenumber $k$ (with oscillations around this curve).

Data from the Planck satellite also appear to show significant deviations from
statistical isotropy in the CMB at very large scales, in the region
$l\lesssim10$ \cite{PlanckXXIII}. As noted by the Planck team, it would be
desirable to have a physically compelling model that provides a common origin
for both the large-scale power deficit and the large-scale anisotropy. We
shall see that our quantum relaxation scenario provides a mechanism for such a
common origin, at least in principle (as was already suggested in ref.
\cite{CV13}). On the basis of numerical simulations of pilot-wave dynamics for
primordial phases, we show that our relaxation scenario naturally predicts
anomalous phases at very large scales as well as a power deficit at comparable scales.

A proper comparison with data is left for future work. In this paper we focus
on delineating the broad features that are to be expected from a quantum
relaxation scenario. We also show by simple estimates that our model is able
to generate a power deficit of approximately the correct magnitude, and at
approximately the correct angular scales, for an appropriate choice of
cosmological parameters. We conclude that our model is potentially viable as
an explanation for both the large-scale power deficit and the large-scale
anisotropy by means of a single mechanism (the suppression of quantum
relaxation at long wavelengths on expanding space).

\section{Background}

In this section we summarise the required background for the implementation of
our model. (For further details see refs. \cite{AV07,AV08,AV10,CV13} and
references therein.)

\subsection{Dynamical suppression of quantum noise at long wavelengths}

In pilot-wave theory, a system has an actual configuration $q(t)$ with a
velocity $\dot{q}\equiv dq/dt$ determined by the wave function $\psi(q,t)$,
where $\psi$ obeys the usual Schr\"{o}dinger equation $i\partial\psi/\partial
t=\hat{H}\psi$ (with $\hbar=1$). For standard Hamiltonians $\hat{H}$, the
velocity $\dot{q}$ is proportional to the gradient $\partial_{q}S$ of the
phase $S$ of $\psi$.\footnote{Historically the theory was proposed in this
form -- with a dynamical law for velocity -- by de Broglie at the 1927 Solvay
conference (for a many-body system) \cite{BV09}. It was revived in a different
form -- with a dynamical law for acceleration, involving a `quantum potential'
-- by Bohm in 1952 \cite{B52a,B52b}. It has recently been shown that Bohm's
version of the dynamics is unstable and therefore untenable \cite{BohmDyn}.}
Quite generally we have $\dot{q}=j/|\psi|^{2}$ where $j=j\left[  \psi\right]
=j(q,t)$ is the Schr\"{o}dinger current \cite{SV08}. In this theory $\psi$ is
a `pilot wave' (defined in configuration space) guiding the motion of an
individual system; it has no \textit{a priori} connection with probabilities.
For an ensemble with the same initial wave function $\psi(q,t_{i})$, it is
possible in pilot-wave theory to consider an arbitrary initial distribution
$\rho(q,t_{i})$ of configurations $q(t_{i})$. The evolving distribution
$\rho(q,t)$ will necessarily satisfy the continuity equation
\begin{equation}
\frac{\partial\rho}{\partial t}+\partial_{q}\cdot\left(  \rho\dot{q}\right)
=0\ .
\end{equation}
Since $\left\vert \psi\right\vert ^{2}$ obeys the same equation (as a simple
consequence of the Schr\"{o}dinger equation), it follows trivially that an
initial `quantum equilibrium' distribution $\rho(q,t_{i})=\left\vert
\psi(q,t_{i})\right\vert ^{2}$ will evolve into a final quantum equilibrium
distribution $\rho(q,t)=\left\vert \psi(q,t)\right\vert ^{2}$. In this
equilibrium state, the probabilities match the Born rule and pilot-wave
dynamics reproduces the empirical predictions of quantum theory
\cite{B52a,B52b}. On the other hand, for an initial nonequilibrium ensemble
($\rho(q,t_{i})\neq\left\vert \psi(q,t_{i})\right\vert ^{2}$) the statistical
predictions will in general disagree with the quantum Born rule. Thus, from a
pilot-wave perspective, quantum physics is a special equilibrium case of a
wider nonequilibrium physics
\cite{AV91a,AV91b,AV92,AV96,AV01,AV02,AV07,AV08,AV09,AV10,AVPwtMw,PV06}.

The quantum-theoretical equilibrium state $\rho_{\mathrm{QT}}=\left\vert
\psi\right\vert ^{2}$ arises from a dynamical process of relaxation (analogous
to thermal relaxation). This process may be quantified by a coarse-grained
$H$-function
\begin{equation}
\bar{H}=\int dq\ \bar{\rho}\ln(\bar{\rho}/\bar{\rho}_{\mathrm{QT}})
\label{Hbar}
\end{equation}
(where $\bar{\rho}$, $\bar{\rho}_{\mathrm{QT}}$ are respectively
coarse-grained values of $\rho$, $\rho_{\mathrm{QT}}$), where $\bar{H}$ obeys
a coarse-graining $H$-theorem $\bar{H}(t)\leq\bar{H}(0)$
\cite{AV91a,AV92,AV01}. The theorem assumes that the initial distributions
have no fine-grained micro-structure. The minimum $\bar{H}=0$ corresponds to
equilibrium ($\bar{\rho}=\bar{\rho}_{\mathrm{QT}}$). Like its classical
analogue, the theorem provides a general mechanism in terms of which one can
understand how equilibrium is approached. The extent to which equilibrium is
actually reached depends on the system and on the initial conditions. For
initial wave functions that are superpositions of energy eigenstates,
numerical simulations demonstrate rapid relaxation $\bar{\rho}\longrightarrow
\bar{\rho}_{\mathrm{QT}}$ on a coarse-grained level
\cite{AV92,AV01,VW05,EC06,TRV12,SC12}, with an approximately exponential decay
of $\bar{H}(t)$ with time \cite{VW05,TRV12}.

Thus we may understand the Born rule as a consequence of a relaxation process
that took place in the remote past, presumably in the very early universe
\cite{AV91a,AV91b,AV92,AV96}. On this basis we may expect ordinary laboratory
systems today -- which have a long and violent astrophysical history -- to
obey the Born rule to high accuracy (in accordance with observation). On the
other hand, initial quantum nonequilibrium could leave observable traces in
the CMB (or perhaps in relic particles that decoupled at sufficiently early
times) \cite{AV01,AV07,AV08,AV09,AV10,CV13}.

To model this process, we consider a spectator scalar field $\phi$ with a
classical Lagrangian density $\mathcal{L}=\frac{1}{2}\sqrt{-g}g^{\mu\nu
}\partial_{\mu}\phi\partial_{\nu}\phi$, evolving on expanding flat space with
line element $d\tau^{2}=dt^{2}-a^{2}d\mathbf{x}^{2}$. Here $a=a(t)$ is the
scale factor and we take $c=1$. We then have
\begin{equation}
\mathcal{L}=\tfrac{1}{2}a^{3}\dot{\phi}^{2}-\tfrac{1}{2}a(\mathbf{\nabla}
\phi)^{2}\ .
\end{equation}
Working in Fourier space and writing the field components as $\phi
_{\mathbf{k}}=\frac{\sqrt{V}}{(2\pi)^{3/2}}\left(  q_{\mathbf{k}
1}+iq_{\mathbf{k}2}\right)  $ -- with real variables $q_{\mathbf{k}r}$ ($r=1$,
$2$) and a normalisation volume $V$ -- the Lagrangian $L=\int d^{3}
\mathbf{x}\;\mathcal{L}$ reads
\[
L=\sum_{\mathbf{k}r}\frac{1}{2}\left(  a^{3}\dot{q}_{\mathbf{k}r}^{2}
-ak^{2}q_{\mathbf{k}r}^{2}\right)  \ .
\]
We then have canonical momenta $\pi_{\mathbf{k}r}\equiv\partial L/\partial
\dot{q}_{\mathbf{k}r}=a^{3}\dot{q}_{\mathbf{k}r}$ and the Hamiltonian becomes
a sum $H=\sum_{\mathbf{k}r}H_{\mathbf{k}r}$ where
\[
H_{\mathbf{k}r}=\frac{1}{2a^{3}}\pi_{\mathbf{k}r}^{2}+\frac{1}{2}
ak^{2}q_{\mathbf{k}r}^{2}
\]
is the Hamiltonian of a harmonic oscillator with time-dependent mass $m=a^{3}$
and time-dependent angular frequency $\omega=k/a$ \cite{AV07,AV08,AV10}. We
focus on the case of a decoupled (that is, unentangled) mode $\mathbf{k}$. If
the wave functional $\Psi$ takes the form $\Psi=\psi_{\mathbf{k}
}(q_{\mathbf{k}1},q_{\mathbf{k}2},t)\varkappa$, where $\varkappa$ depends only
on degrees of freedom for modes $\mathbf{k}^{\prime}\neq\mathbf{k}$, we obtain
an independent dynamics for the mode with wave function $\psi_{\mathbf{k}
}(q_{\mathbf{k}1},q_{\mathbf{k}2},t)$.

Dropping the index $\mathbf{k}$, the wave function $\psi=\psi(q_{1},q_{2},t)$
satisfies the Schr\"{o}dinger equation
\begin{equation}
i\frac{\partial\psi}{\partial t}=\sum_{r=1,\ 2}\left(  -\frac{1}{2m}
\partial_{r}^{2}+\frac{1}{2}m\omega^{2}q_{r}^{2}\right)  \psi\ , \label{S2D}
\end{equation}
while de Broglie's equation of motion for the configuration $(q_{1},q_{2})$
reads
\begin{equation}
\dot{q}_{r}=\frac{1}{m}\operatorname{Im}\frac{\partial_{r}\psi}{\psi}
\label{deB2D}
\end{equation}
(with $\partial_{r}\equiv\partial/\partial q_{r}$). The marginal distribution
$\rho=\rho(q_{1},q_{2},t)$ for the mode evolves according to the continuity
equation
\begin{equation}
\frac{\partial\rho}{\partial t}+\sum_{r=1,\ 2}\partial_{r}\left(  \rho\frac
{1}{m}\operatorname{Im}\frac{\partial_{r}\psi}{\psi}\right)  =0\ .
\label{CE2D}
\end{equation}
Thus we may discuss relaxation for a single field mode in terms of relaxation
for a harmonic oscillator with time-dependent mass and angular frequency
\cite{AV07,AV08}.

We study the case of a radiation-dominated expansion, $a\propto t^{1/2}$. We
consider that our results model a relaxation process taking place during a
pre-inflationary era. The field $\phi$ is taken to model the behaviour of
whatever generic fields may have been present at that time. The relation
between our field $\phi$ and particular fields such as the inflaton field is
not really known or specified, pending the future development of a more
detailed model. Our aim here is to obtain general features that could emerge
from an incomplete relaxation to quantum equilibrium during pre-inflation.

It should be noted that, in what follows, equation (\ref{CE2D}) does all of
the mathematical work in generating the results. This same equation appears in
standard quantum theory as a simple consequence of the Schr\"{o}dinger
equation. The key difference is that here we allow ourselves to evolve this
equation forward in time starting from anomalous initial conditions that
violate the Born rule -- a possibility that makes no sense in ordinary quantum
theory but which is perfectly natural in pilot-wave theory. Specifically, we
assume that at the initial time the width of $\rho(q_{1},q_{2},t_{i})$ is
smaller than the width of $|\psi(q_{1},q_{2},t_{i})|^{2}$. This
(mathematically) tiny change might provide a common origin for the observed
large-scale cosmic anomalies.

It has been shown that the time evolution of our field mode on expanding
space, as defined by equations (\ref{S2D})--(\ref{CE2D}), is mathematically
equivalent to the time evolution of a standard harmonic oscillator with real
time $t$ replaced by a `retarded time' $t_{\mathrm{ret}}(t)$ that depends on
the wavenumber $k$ of the mode \cite{CV13}. (The equivalence also requires the
use of appropriately rescaled variables for each system.)

Defining a parameter
\begin{equation}
\varepsilon\equiv\left(  \frac{t_{i}}{a_{i}^{2}}\right)  k^{2}\ ,
\label{epsilon}
\end{equation}
for completeness we note that the retarded time is given by
\begin{equation}
t_{\mathrm{ret}}(t)=t_{i}+\frac{1}{\omega_{i}}\Theta(t)
\end{equation}
where $\omega_{i}$ is equal to $\omega$ at time $t_{i}$ and where (for
$a=a_{i}(t/t_{i})^{1/2}$)

\begin{align}
\Theta(t)  &  =\tan^{-1}\left(  {\frac{1+4\varepsilon t_{i}}{4\varepsilon
t_{i}}\tan{(2\sqrt{\varepsilon t}-2\sqrt{\varepsilon t_{i}})}+\frac{1}
{2\sqrt{\varepsilon t_{i}}}}\right) \nonumber\\
&  +\pi.\text{nint}(\frac{2\sqrt{\varepsilon t}-2\sqrt{\varepsilon t_{i}}}
{\pi})-\tan^{-1}\left(  \frac{1}{2\sqrt{\varepsilon t_{i}}}\right)  
\label{Thetaeval}
\end{align}
(with $\text{nint}(x)$ returning the integer nearest to $x$) \cite{CV13}.

This result provides us with a convenient means of performing simulations. A
desired time evolution from initial conditions at $t_{i}$ to final conditions
at $t_{f}$ may be obtained by evolving a standard harmonic oscillator (with
the same initial conditions) from $t_{i}$ to $t_{\mathrm{ret}}(t_{f})$. We
emphasise, however, that this is simply a convenient means of evolving the
continuity equation (\ref{CE2D}) forwards in time for a field mode on
expanding space. One could simply integrate this equation directly; the
results will be the same.

In the short-wavelength (sub-Hubble) limit $t_{\mathrm{ret}}(t)$ reduces to
real time $t$ and we recover the evolution of a field mode on Minkowski
spacetime. In this limit, for a superposition of excited states, relaxation
will take place rapidly as for an ordinary oscillator. In contrast, at long
(super-Hubble) wavelengths $t_{\mathrm{ret}}(t)<<t$ and relaxation is
retarded. (For a detailed discussion see ref. \cite{CV13}.) Relaxation
suppression at super-Hubble wavelengths may also be understood in terms of an
upper bound on the mean displacement of the trajectories \cite{AV08,AVbook}.

Let us consider an initial wave function that is a superposition
\begin{equation}
\psi(q_{1},q_{2},t_{i})=\frac{1}{\sqrt{M}}\sum_{n_{1}=0}^{\sqrt{M}-1}
\sum_{n_{2}=0}^{\sqrt{M}-1}e^{i\theta_{n_{1}n_{2}}}\Phi_{n_{1}}(q_{1}
)\Phi_{n_{2}}(q_{2}) \label{psiinitial}
\end{equation}
of $M$ energy eigenstates $\Phi_{n_{1}}\Phi_{n_{2}}$ of the initial
Hamiltonian. The coefficients $c_{n_{1}n_{2}}(t_{i})=(1/\sqrt{M}
)e^{i\theta_{n_{1}n_{2}}}$ have equal amplitude and randomly-chosen phases
$\theta_{n_{1}n_{2}}$. (Because $n_{1}$, $n_{2}$ have the same range the
number $M$ of modes is the square of an integer.) The wave function at time
$t$ is then
\begin{equation}
\psi(q_{1},q_{2},t)=\frac{1}{\sqrt{M}}\sum_{n_{1}=0}^{\sqrt{M}-1}\sum
_{n_{2}=0}^{\sqrt{M}-1}e^{i\theta_{n_{1}n_{2}}}\psi_{n_{1}}(q_{1}
,t)\psi_{n_{2}}(q_{2},t)\ , \label{psit}
\end{equation}
where the exact solution for $\psi_{n}(q,t)$ is given by equation (19) of ref.
\cite{CV13}. At time $t$ we have an equilibrium distribution $\rho
_{\mathrm{QT}}(q_{1},q_{2},t)=|\psi(q_{1},q_{2},t)|^{2}$. As in ref.
\cite{CV13}, we take an initial nonequilibrium distribution
\begin{equation}
\rho(q_{1},q_{2},t_{i})=|\Phi_{0}(q_{1})\Phi_{0}(q_{2})|^{2}=\frac{\omega
_{i}m_{i}}{\pi}e^{-m_{i}\omega_{i}q_{1}^{2}}e^{-m_{i}\omega_{i}q_{2}^{2}}
\label{rhoinitial}
\end{equation}
(equal to the equilibrium distribution for the ground state $\Phi_{0}
(q_{1})\Phi_{0}(q_{2})$). This is chosen on grounds of simplicity only.
Clearly $\rho(q_{1},q_{2},t_{i})\neq|\psi(q_{1},q_{2},t_{i})|^{2}$ and the
initial width (or variance)\ is smaller than the equilibrium width. By
calculating the time evolution $\rho(q_{1},q_{2},t)$ of the ensemble
distribution one may study the extent to which it approaches the equilibrium
distribution $|\psi(q_{1},q_{2},t)|^{2}$ (on a coarse-grained level).

In our simulations the time evolution of $\rho$ is reconstructed from a
calculation of trajectories traversing a fine grid, where the trajectories are
simulated using the equivalence to a standard oscillator with a retarded time.
(For details see ref. \cite{CV13}.) But again these are merely convenient
techniques for evolving (\ref{CE2D}) forwards in time. As we have noted, our
results follow from equation (\ref{CE2D}) alone.

In ref. \cite{CV13} we performed an illustrative numerical simulation of the
evolution of $\rho(q_{1},q_{2},t)$ for the case of $M=25$ energy states. We
considered a field mode of wavenumber $k$ such that the initial (physical)
wavelength was ten times the initial Hubble radius and we evolved forwards to
a final time $t_{f}=t_{\mathrm{enter}}(k)$ where $t_{\mathrm{enter}}(k)$ is
the time of mode entry. This example served to illustrate time evolution in
the super-Hubble regime. Only a partial relaxation towards equilibrium was
observed. In particular, the support of $\rho$ remained significantly narrower
than the support of $|\psi|^{2}$, the final width of the former being about
one half of the final width of the latter (see Figure 2 of ref. \cite{CV13}).
Whereas with no expanding space -- or equivalently, in the short-wavelength
(Minkowski) limit -- over the same period of time there is almost complete
relaxation: the final distributions $\rho$ and $|\psi|^{2}$ match very closely
(on a coarse-grained level) as regards both detailed features and their
respective widths (see Figure 3 of ref. \cite{CV13}).

The contrast between these results illustrates the retardation or suppression
of relaxation in the super-Hubble regime as compared with the short-wavelength limit.

\subsection{Primordial quantum nonequilibrium and the CMB}

Such suppression of relaxation may have occurred during a radiation-dominated
pre-inflationary era \cite{AV07,AV08,AV10,CV13}. It is of particular interest
to consider a pre-inflationary phase with a small number of excitations above
the vacuum, since it is generally assumed that during inflation itself the
quantum state is in or very close to the vacuum. For the two-dimensional
harmonic oscillator it has been found that even for a quantum state with a
minimal number of excitations, of the form $\sim\left\vert 00\right\rangle
+\left\vert 01\right\rangle +\left\vert 10\right\rangle +\left\vert
11\right\rangle $ (with randomly-chosen relative initial phases), relaxation
still takes place (at least to a good approximation) over sufficiently long
timescales \cite{LongRel}. Therefore, if the pre-inflationary phase lasts long
enough, at the onset of inflation we can expect approximate equilibrium at
short wavelengths even with a tiny number ($M=4$) of excited pre-inflationary
states. This is an important feature because a significant number of
excitations above the inflationary vacuum is likely to cause a back-reaction
problem \cite{BM13}.\footnote{AV is grateful to J\'{e}r\^{o}me Martin for
helpful discussions of this point.}

According to our proposed scenario, the spectrum of perturbations that remains
at the end of the pre-inflationary era will seed the spectrum at the onset of
inflation. It has been shown that, during inflation itself, no further
relaxation takes place and the perturbations are simply transferred to larger
lengthscales \cite{AV07,AV10}. By this means, incomplete relaxation during the
pre-inflationary era can affect the spectrum of primordial perturbations that
generate the temperature anisotropy in the CMB (and that trigger the formation
of large-scale structure generally).

Implicit in this scenario is the assumption that the spectrum at the end of
pre-inflation will survive the transition to inflation and seed the
inflationary spectrum. The assumption seems plausible, since we have a
transition from pre-inflation with relaxation suppression on super-Hubble
scales to inflation with completely suppressed relaxation on all scales. It
then appears possible that super-Hubble modes that are out of equilibrium just
before the transition will not completely relax during the transition.
However, to test this assumption requires a model of the transition and a
study of how nonequilibrium modes will evolve across the transition. This is
left for future work. Strictly speaking, the results presented in this paper
are for the spectrum at the end of a radiation-dominated era. To apply our
results as a possible explanation for the observed large-scale cosmic
anomalies requires us to assume that the spectrum will not be greatly affected
by the transition.\footnote{One might suggest a simpler scenario in which
initial nonequilibrium conditions are set at the beginning of inflation
itself, thereby avoiding the complication of a transition from pre-inflation
to inflation. However, because there is no relaxation during inflation (on all
scales) the final correction to the power spectrum would be simply equal to
the correction that is assumed at the beginning. In such a scenario one can
use observations to set bounds on the initial nonequilibrium (as studied in
ref. \cite{AV10}) -- but one cannot make predictions. In the present paper we
obtain predictions that stem from wavelength-dependent relaxation on a
radiation-dominated pre-inflationary background (as suggested in refs.
\cite{AV10,CV13}). This requires that we set our initial nonequilibrium
conditions at the beginning of the pre-inflationary era.}

Note also that pre-inflationary nonequilibrium super-Hubble modes can
contribute to the CMB spectrum only if they are driven inside the Hubble
radius during the transition to inflation. This requires that the comoving
Hubble radius increases during the transition. It was shown in ref.
\cite{CV13} that this can occur for a reasonable time variation of the
equation-of-state parameter.

We emphasise that we study a spectator scalar field, without a specific
inflationary model. We do not know how this field is related to the inflaton
field or to other perturbative fields. We take the behaviour of our field as a
simple model of presumably generic field behaviour, and we assume that broad
features of its spectrum will be similarly present for the relevant fields in
a full model. The complexity of our numerical simulations necessarily
restricts us to a simplified model, at least at this initial exploratory
stage. We hope in future work to develop more complete models.

With these assumptions and caveats, we may tentatively apply our results to
the possible interpretation of large-scale cosmic anomalies.

Before proceeding, let us briefly recap how quantum nonequilibrium can
generate corrections to predictions for the CMB \cite{AV07,AV10}.

As we have noted, there is no relaxation during the inflationary era itself.
An inflaton perturbation $\phi_{\mathbf{k}}$ generates a curvature
perturbation $\mathcal{R}_{\mathbf{k}}\propto\phi_{\mathbf{k}}$ (more
precisely, $\mathcal{R}_{\mathbf{k}}$ is proportional to the late-time
perturbation $\phi_{\mathbf{k}}$ evaluated at a time a few e-folds after the
mode exits the Hubble radius) \cite{LL00}. This in turn generates coefficients
\cite{LR99}
\begin{equation}
a_{lm}=\frac{i^{l}}{2\pi^{2}}\int d^{3}\mathbf{k}\ \mathcal{T}(k,l)\mathcal{R}
_{\mathbf{k}}Y_{lm}(\mathbf{\hat{k}}) \label{alm}
\end{equation}
(where $\mathcal{T}(k,l)$ is the transfer function) that appear in the
spherical harmonic expansion
\begin{equation}
\frac{\Delta T(\theta,\phi)}{\bar{T}}=\sum_{l=2}^{\infty}\sum_{m=-l}
^{+l}a_{lm}Y_{lm}(\theta,\phi) \label{har}
\end{equation}
of the observed CMB temperature anisotropy. Statistical isotropy for $\Delta
T(\theta,\phi)$ implies that
\begin{equation}
\left\langle a_{l^{\prime}m^{\prime}}^{\ast}a_{lm}\right\rangle =\delta
_{ll^{\prime}}\delta_{mm^{\prime}}C_{l}\ , \label{iso}
\end{equation}
where $\left\langle ...\right\rangle $ denotes an average over the underlying
theoretical ensemble and $C_{l}\equiv\left\langle \left\vert a_{lm}\right\vert
^{2}\right\rangle $ is the angular power spectrum \cite{Muk05,V14}.
Statistical homogeneity for $\mathcal{R}_{\mathbf{k}}$ implies further that
$\left\langle \mathcal{R}_{\mathbf{k}}\mathcal{R}_{\mathbf{k
\acute{}}}^{\ast}\right\rangle =\delta_{\mathbf{kk}\acute{}
}\left\langle \left\vert \mathcal{R}_{\mathbf{k}}\right\vert ^{2}\right\rangle
$. From (\ref{alm}) we then have
\begin{equation}
C_{l}=\frac{1}{2\pi^{2}}\int_{0}^{\infty}\frac{dk}{k}\ \mathcal{T}
^{2}(k,l)\mathcal{P}_{\mathcal{R}}(k)\ , \label{Cl2}
\end{equation}
where%
\begin{equation}
\mathcal{P}_{\mathcal{R}}(k)\equiv\frac{4\pi k^{3}}{V}\left\langle \left\vert
\mathcal{R}_{\mathbf{k}}\right\vert ^{2}\right\rangle \label{PPS}%
\end{equation}
is the primordial power spectrum (with $V$ a normalisation volume).

Thus, observational constraints on $C_{l}$ imply observational constraints on
$\mathcal{P}_{\mathcal{R}}(k)$ and hence (since $\mathcal{R}_{\mathbf{k}%
}\propto\phi_{\mathbf{k}}$) observational constraints on the primordial
variance $\left\langle \left\vert \phi_{\mathbf{k}}\right\vert ^{2}%
\right\rangle $ for $\phi_{\mathbf{k}}$. Writing%
\begin{equation}
\left\langle |\phi_{\mathbf{k}}|^{2}\right\rangle =\left\langle |\phi
_{\mathbf{k}}|^{2}\right\rangle _{\mathrm{QT}}\xi(k)\ , \label{xi}%
\end{equation}
where $\left\langle ...\right\rangle _{\mathrm{QT}}$ denotes the
quantum-theoretical expectation value, we have%
\begin{equation}
\mathcal{P}_{\mathcal{R}}(k)=\mathcal{P}_{\mathcal{R}}^{\mathrm{QT}}(k)\xi(k)
\label{xi2}%
\end{equation}
where $\mathcal{P}_{\mathcal{R}}^{\mathrm{QT}}(k)$ is the primordial power
spectrum predicted by quantum theory. Measurements of the angular power
spectrum $C_{l}$ may then be used to set experimental bounds on $\xi(k)$
\cite{AV10}.

The `nonequilibrium function' $\xi(k)$ measures the power deficit (if
$\xi(k)<1$) as a function of $k$. We expect $\xi(k)$ to be smaller for smaller
$k$ -- since during pre-inflation there will be more suppression of relaxation
at longer wavelengths -- while we expect $\xi(k)$ to approach $1$ in the
short-wavelength limit of large $k$. But can we make a precise
\textit{prediction} for $\xi(k)$ as a function of $k$? It would be of interest
to obtain quantitative predictions for the shape of the curve $\xi=\xi(k)$ and
to compare these with the data for $\mathcal{P}_{\mathcal{R}}(k)$ and $C_{l}$.

\section{Predictions for the power deficit}

To obtain a prediction for the deficit function $\xi=\xi(k)$, we must repeat
the simulation of ref. \cite{CV13} for varying values of $k$, calculate
$\xi(k)$ for each and plot the results (as a function of $k$). We should also
repeat the simulations for varying numbers $M$ of energy states and for
varying final times $t_{f}$, with a view to finding features of the function
$\xi(k)$ that are as far as possible independent of details of the
pre-inflationary era -- features that might provide an observational signature
of primordial quantum nonequilibrium (as opposed to a mere generic power
deficit that could equally be produced by other effects).

For a given pre-inflationary wave function $\psi(q_{1},q_{2},t)$ and
distribution $\rho(q_{1},q_{2},t)$, each degree of freedom $q_{r}$ has an
equilibrium variance $\Delta_{r}^{2}=\left\langle q_{r}^{2}\right\rangle
_{\mathrm{QT}}-\left\langle q_{r}\right\rangle _{\mathrm{QT}}^{2}$ and a
nonequilibrium variance $D_{r}^{2}=\left\langle q_{r}^{2}\right\rangle
-\left\langle q_{r}\right\rangle ^{2}$ -- where $\left\langle ...\right\rangle
_{\mathrm{QT}}$ and $\left\langle ...\right\rangle $ denote averaging with
respect to $\left\vert \psi(q_{1},q_{2},t)\right\vert ^{2}$ and $\rho
(q_{1},q_{2},t)$ respectively. Equation (\ref{xi}) defines $\xi(k)$ as the
ratio%
\[
\frac{\left\langle |\phi_{\mathbf{k}}|^{2}\right\rangle }{\left\langle
|\phi_{\mathbf{k}}|^{2}\right\rangle _{\mathrm{QT}}}=\frac{\left\langle
q_{1}^{2}\right\rangle +\left\langle q_{2}^{2}\right\rangle }{\left\langle
q_{1}^{2}\right\rangle _{\mathrm{QT}}+\left\langle q_{2}^{2}\right\rangle
_{\mathrm{QT}}}\ ,
\]
where $\phi_{\mathbf{k}}$ is the inflaton perturbation defined during the
inflationary era. In our pre-inflationary model, in contrast, we shall take
$\xi(k)$ to be defined by%
\begin{equation}
\xi(k)=\frac{D_{1}^{2}+D_{2}^{2}}{\Delta_{1}^{2}+\Delta_{2}^{2}}%
=\frac{\left\langle q_{1}^{2}\right\rangle -\left\langle q_{1}\right\rangle
^{2}+\left\langle q_{2}^{2}\right\rangle -\left\langle q_{2}\right\rangle
^{2}}{\left\langle q_{1}^{2}\right\rangle _{\mathrm{QT}}-\left\langle
q_{1}\right\rangle _{\mathrm{QT}}^{2}+\left\langle q_{2}^{2}\right\rangle
_{\mathrm{QT}}-\left\langle q_{2}\right\rangle _{\mathrm{QT}}^{2}}\ ,
\label{pureksi}%
\end{equation}
where $\phi_{\mathbf{k}}$ is our pre-inflationary scalar field. The reason for
adopting the definition (\ref{pureksi}) is that, if our pre-inflationary
spectrum is to act as a seed for perturbations at the beginning of inflation
then the mean values should be subtracted. Thus we may define effective
pre-inflationary perturbations $\tilde{\phi}_{\mathbf{k}}\equiv\phi
_{\mathbf{k}}-\left\langle \phi_{\mathbf{k}}\right\rangle $ and $\tilde{\phi
}_{\mathbf{k}}^{\mathrm{QT}}\equiv\phi_{\mathbf{k}}-\left\langle
\phi_{\mathbf{k}}\right\rangle _{\mathrm{QT}}$ for the respective
nonequilibrium and equilibrium cases (so that $\left\langle \left\vert
\tilde{\phi}_{\mathbf{k}}\right\vert ^{2}\right\rangle =\left\langle
|\phi_{\mathbf{k}}|^{2}\right\rangle -|\left\langle \phi_{\mathbf{k}%
}\right\rangle |^{2}$ and $\left\langle \left\vert \tilde{\phi}_{\mathbf{k}%
}^{\mathrm{QT}}\right\vert ^{2}\right\rangle _{\mathrm{QT}}=\left\langle
|\phi_{\mathbf{k}}|^{2}\right\rangle _{\mathrm{QT}}-|\left\langle
\phi_{\mathbf{k}}\right\rangle _{\mathrm{QT}}|^{2}$). We may then take
$\xi(k)=\left\langle \left\vert \tilde{\phi}_{\mathbf{k}}\right\vert
^{2}\right\rangle /\left\langle \left\vert \tilde{\phi}_{\mathbf{k}%
}^{\mathrm{QT}}\right\vert ^{2}\right\rangle _{\mathrm{QT}}$, which is equal
to (\ref{pureksi}). In effect, the definition (\ref{pureksi}) subtracts the
mean values of the pre-inflationary perturbations.

So far we have defined $\xi(k)$ for a pure quantum state. In general we would
expect the pre-inflationary era to be in a mixed quantum state. A decoupled
mode $\mathbf{k}$ will have a density operator%
\[
\hat{\rho}=\sum_{n}p_{n}\left\vert \psi_{n}\right\rangle \left\langle \psi
_{n}\right\vert \ ,
\]
where the wave functions $\psi_{n}=\psi_{n}(q_{1},q_{2},t)$ are distinct
superpositions (with different numbers $M$ of modes, and coefficients with
different amplitudes and phases).\footnote{Note that in pilot-wave theory a
mixed quantum state is interpreted in terms of a preferred decomposition.} The
observed or effective function $\xi(k)$ will then be obtained by appropriate
averaging over the statistical mixture of $\psi_{n}$'s.

Consider a given wave number $k$. For each $\psi_{n}$ we may evolve the
initial nonequilibrium distribution (\ref{rhoinitial}) forwards in time (from
$t_{i}$ to $t_{f}$) to obtain a final distribution $\rho_{n}(q_{1},q_{2}%
,t_{f})$. We may then calculate the final variances $\Delta_{rn}^{2}(k)$ and
$D_{rn}^{2}(k)$ of $q_{r}$ for the respective distributions $|\psi_{n}%
(q_{1},q_{2},t_{f})|^{2}$ and $\rho_{n}(q_{1},q_{2},t_{f})$. For a mixed state
we take $\xi(k)$ to be defined by%
\begin{equation}
\xi(k)=\frac{\left\langle D_{1n}^{2}+D_{2n}^{2}\right\rangle _{\mathrm{mixed}%
}}{\left\langle \Delta_{1n}^{2}+\Delta_{2n}^{2}\right\rangle _{\mathrm{mixed}%
}}\ , \label{xi_mixed}%
\end{equation}
where $\left\langle ...\right\rangle _{\mathrm{mixed}}$ denotes a statistical
average over the mixture of $\psi_{n}$'s.

For simplicity we focus on mixtures whose component wave functions take the
form (\ref{psit}), with a fixed number $M$ of modes with equally-weighted
amplitudes but with randomly-chosen initial phases $\theta_{n_{1}n_{2}}%
$.\textbf{ }The mean $\left\langle ...\right\rangle _{\mathrm{mixed}}$ then
amounts to an average over different sets of initial phases, where the index
$n$ now labels the set of initial phases that characterises the quantum state
$\psi_{n}$. (It would also be of interest to study mixtures of wave functions
with different values of $M$ but we leave this for future work.)

Such calculations are computationally intensive. As in ref. \cite{CV13}, we
evolve over a fixed time interval $(t_{i},t_{f})=(10^{-4},10^{-2})$ (with
units $\hslash=c=1$), where for convenience we take $a_{0}=1$ at $t_{0}=1$.
The calculation is performed for varying values of $k$ and $M$, keeping the
time interval fixed. Later, we also look at varying $t_{f}$ for fixed
$M$.\footnote{In ref. \cite{CV13} we employed a fifth-order Runge-Kutta method
(due to Dormand and Prince and often denoted DOPRI5). Here we employ an
eighth-order Runge-Kutta method with a more robust error estimation (as
developed by Dormand and Prince, refined by Hairer, N\o rsett and Wanner, and
often denoted DOPRI853) \cite{HNW1993,NR2007}.}

Note that at the final time $t_{f}=10^{-2}$ we have a scale factor
$a_{f}=t_{f}^{1/2}=0.1$ and a Hubble radius $H_{f}^{-1}=2t_{f}=0.02$. For the
mode with $\lambda=0.2$ or $k=2\pi/\lambda=10\pi$, the final physical
wavelength $\lambda_{\mathrm{phys}}(t_{f})=a_{f}\lambda=0.02$ is equal to the
final Hubble radius $H_{f}^{-1}$. For smaller $k$ ($<10\pi$) the mode will be
outside the Hubble radius at $t_{f}$; for larger $k$ ($>10\pi$) the mode will
be inside the Hubble radius at $t_{f}$.

In our simulations we use natural units with $\hslash=c=1$, in which time has
dimensions of an inverse mass. With $\hbar\sim10^{-33}\ \mathrm{Js}$, an
initial time $10^{-37}\ \mathrm{s}$ in standard units corresponds to an
initial time $10^{-37}/\hbar\sim10^{-4}$ in our units. These numbers are not
intended to have any special significance; they are chosen for numerical
convenience only.

We proceed as follows. For each $M$ the calculation is repeated for varying
values of $k$.\textbf{ }For each $k$ six separate calculations are performed
with different sets of randomly-chosen initial phases, yielding results for
$D_{1n}^{2}+D_{2n}^{2}$ and $\Delta_{1n}^{2}+\Delta_{2n}^{2}$ with six
different values of $n$. We then calculate the averages $\left\langle
D_{1n}^{2}+D_{2n}^{2}\right\rangle _{\mathrm{mixed}}$ and $\left\langle
\Delta_{1n}^{2}+\Delta_{2n}^{2}\right\rangle _{\mathrm{mixed}}$ over these six
results and thus obtain an estimate for the ratio (\ref{xi_mixed}).

\begin{figure}
\begin{center}
\includegraphics[width=0.9\textwidth]{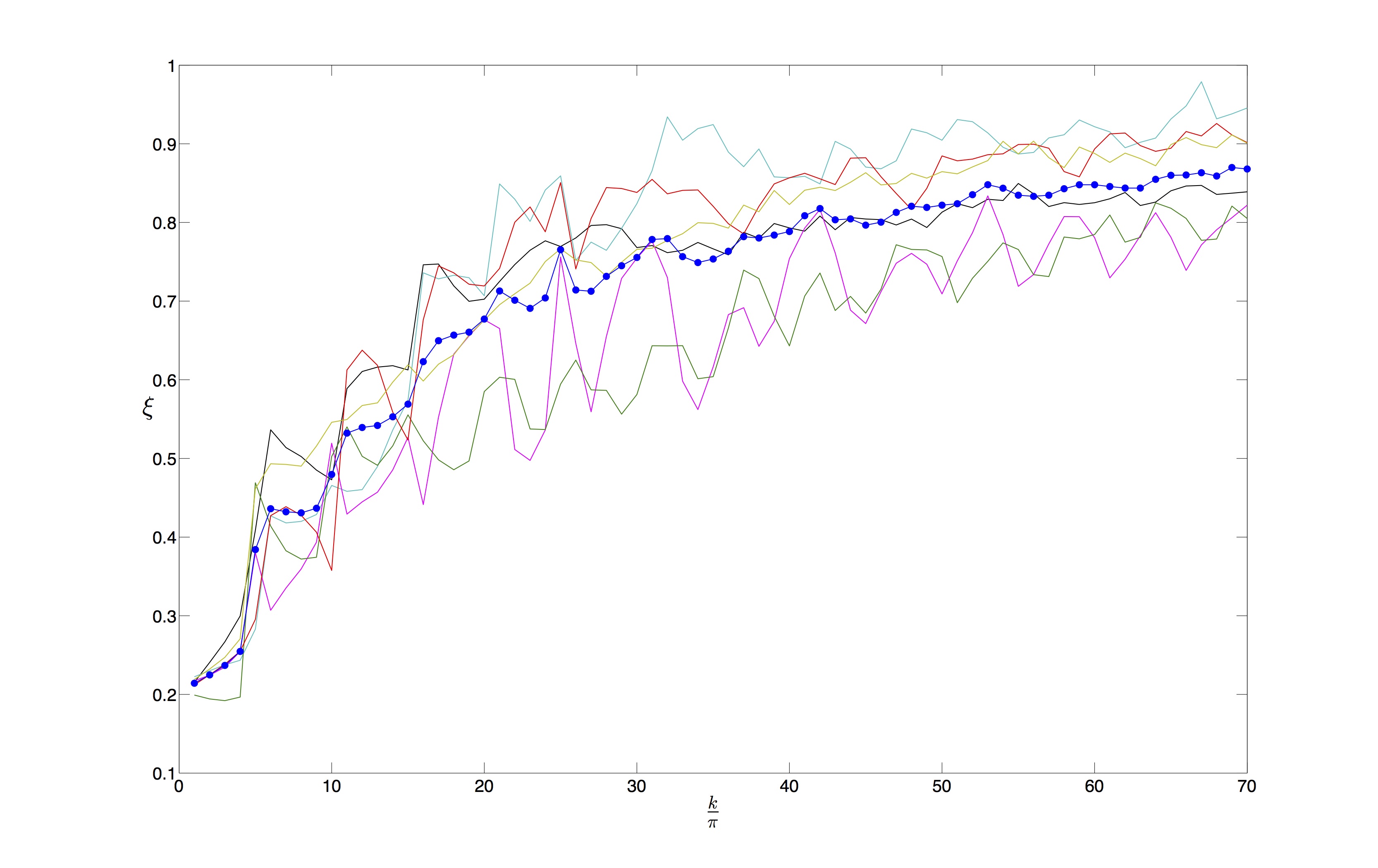}
\caption{\label{fig1}Simulations for $M=25$ modes, showing the mixed-ensemble curve
$\xi=\xi(k)$ (in blue with bullets). For comparison we also plot the six
`pure-ensemble' curves $\xi_{n}=\xi_{n}(k)$ each obtained from the ratio
$(D_{1n}^{2}+D_{2n}^{2})/(\Delta_{1n}^{2}+\Delta_{2n}^{2})$ for a single set
of (randomly-chosen) initial phases. The mixed-ensemble curve $\xi(k)$ is more
relevant to observation.}\end{center}
\end{figure}

An example is shown in Figure \ref{fig1}  for $M=25$. We plot results for $k=n\pi$ with $n=1,2,3,...,70$. We have found it difficult to calculate
accurately beyond $k=80\pi$ where the normalisation of the density starts to
deviate significantly from unity, since the number of inaccurate trajectories
is too high. In addition to the mixed-ensemble curve $\xi(k)$ -- shown in blue
with bullets -- for comparison we also display six `pure-ensemble' curves
$\xi_{n}(k)$ each obtained from the values of $(D_{1n}^{2}+D_{2n}^{2}%
)/(\Delta_{1n}^{2}+\Delta_{2n}^{2})$ for a single $n$ (that is, for a single
set of initial phases). The curves $\xi_{n}(k)$ show rather large
oscillations. The curve $\xi(k)$ is considerably smoother but still shows
oscillations, though these appear to be damped for larger $k$.\footnote{Note
that $\xi(k)$ differs from the ensemble mean $\left\langle \xi_{n}%
(k)\right\rangle _{\mathrm{mixed}}$ of the $\xi_{n}(k)$'s. In our definition
(\ref{xi_mixed}) we calculate the ensemble averages of the variances and then
take the ratio of the results. Whereas for $\left\langle \xi_{n}%
(k)\right\rangle _{\mathrm{mixed}}$ the ratio of the variances would be taken
for each $n$ before averaging over the ensemble. In practice we find that
numerically there is not much difference between $\xi(k)$ and $\left\langle
\xi_{n}(k)\right\rangle _{\mathrm{mixed}}$. However, strictly speaking
$\xi(k)$ is the physically relevant quantity.}%

\subsection{Fixed time interval and varying number of modes}

\begin{figure}
\begin{center}
\includegraphics[width=0.9\textwidth]{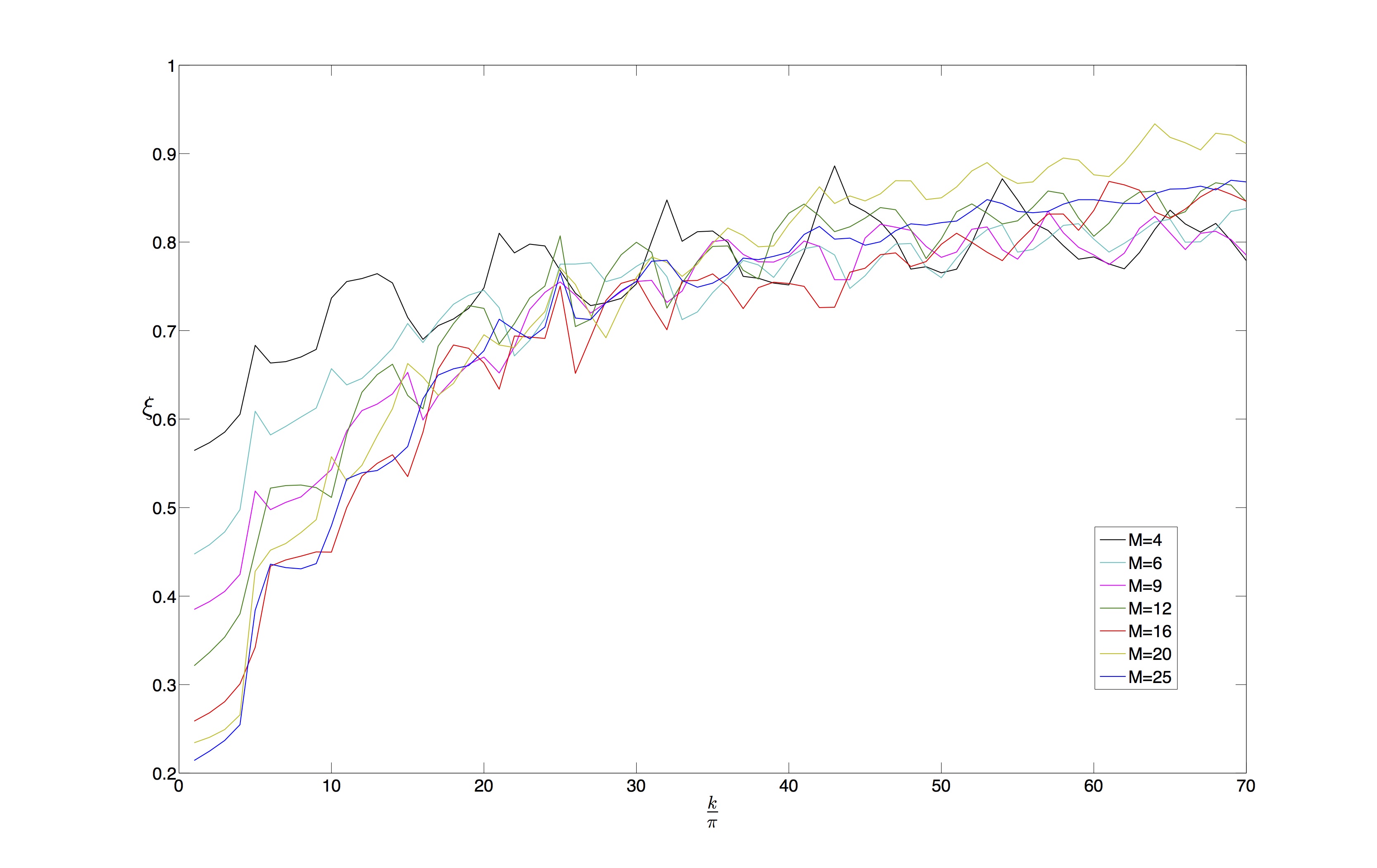}
\caption{\label{fig2}Mixed-ensemble curves $\xi(k)$ for varying $M=4,6,9,12,16,20,25$ with
a fixed time interval $(t_{i},t_{f})=(10^{-4},10^{-2})$.}
\end{center}
\end{figure}

We first consider the mixed-ensemble curve $\xi=\xi(k)$ obtained from
evolution over a fixed time interval $(t_{i},t_{f})=(10^{-4},10^{-2})$ and for
varying values of $M$. The results are shown in Figure \ref{fig2}, for
$M=4,6,9,12,16,20,25$. The curves show some interesting small-scale features.
But to a first approximation we may focus on the smooth, overall structure and
try fitting to a simple function with no oscillations. (Fits that include
oscillations will be considered in Section 3.3.)

For each $M$ we find a best fit of $\xi(k)$ to the curve
\begin{equation}
\xi(k)=\tan^{-1}(c_{1}\frac{k}{\pi}+c_{2})-\frac{\pi}{2}+c_{3} \label{fit1}
\end{equation}
where $c_{1}$, $c_{2}$ and $c_{3}$ are free parameters.

Note that $\xi\rightarrow c_{3}$ as $k\rightarrow\infty$. Thus this choice of
fitting function includes the possibility that $\xi$ does not approach unity
at large $k$ -- in which case there will be a residual nonequilibrium $\xi<1$
even in the short-wavelength limit. (For example, for $M=12$ we will find that
$c_{3}=0.95$.) Such a residue would in effect induce an overall
renormalisation of the observed power spectrum and would therefore in itself
not really be observable -- or at least not distinguishable from an equivalent
shift in other cosmological parameters. For example, in standard models the
primordial power spectrum $\mathcal{P}_{\mathcal{R}}^{\mathrm{QT}}(k)$ is to
lowest-order proportional to the fourth power of the Hubble parameter
$H_{\mathrm{\inf}}$ during inflation, and this parameter is subject to a large
uncertainty. We shall see that our residue $c_{3}$ is only slightly less than
$1$ and so could easily be offset by a small increase in $H_{\inf}$.

Given the best fits for each $M$, we may study how the parameters $c_{1}$,
$c_{2}$ and $c_{3}$ depend on $M$, in the hope of extracting general features.

From Figure \ref{fig2} we may discern some overall trends: (i) At low $k$ we see that
$\xi$ is smaller for larger $M$. This is simply because at low $k$ little
evolution has taken place (owing to retardation at long wavelengths) and so
the distributions approximate their initial values, where for larger $M$ the
initial equilibrium distribution has a larger spread. (ii) There are
oscillations in $\xi(k)$. While the oscillations appear regular in some cases
(notably $M=4$ and $16$) for others they are rather erratic (for example
$M=9$). (iii) At high $k$ there is an approximate convergence of $\xi$ towards
$1$.

As one would expect, $\xi$ generally reaches closer to $1$ for larger $M$.
(This is expected since for a given time interval there will be more
relaxation for larger $M$.) As we shall see presently, the best-fit limiting
value $c_{3}$ generally increases for increasing $M$. However, the curve for
$M=20$ is in this respect somewhat puzzling, since the corresponding value of
$c_{3}$ is found to be significantly larger than for $M=25$ in contradiction
with the overall trend. This is clear by eye from Figure \ref{fig2}. The curve for
$M=20$ begins approximately mid-way between the curves for $M=16$ and $M=25$,
and yet it ends significantly higher than any of the other curves. At present
we have no explanation for this seemingly anomalous result for $M=20$. In
attempting to extract a general functional dependence for the parameters
$c_{1}$, $c_{2}$ and $c_{3}$, we find it convenient to omit the results for
$M=20$. Pending further understanding, it seems reasonable to discount this
case -- especially since, as we shall see, taken on their own the other
results mostly follow a clear and simple pattern.

\begin{figure}
\begin{center}
\includegraphics[width=0.9\textwidth]{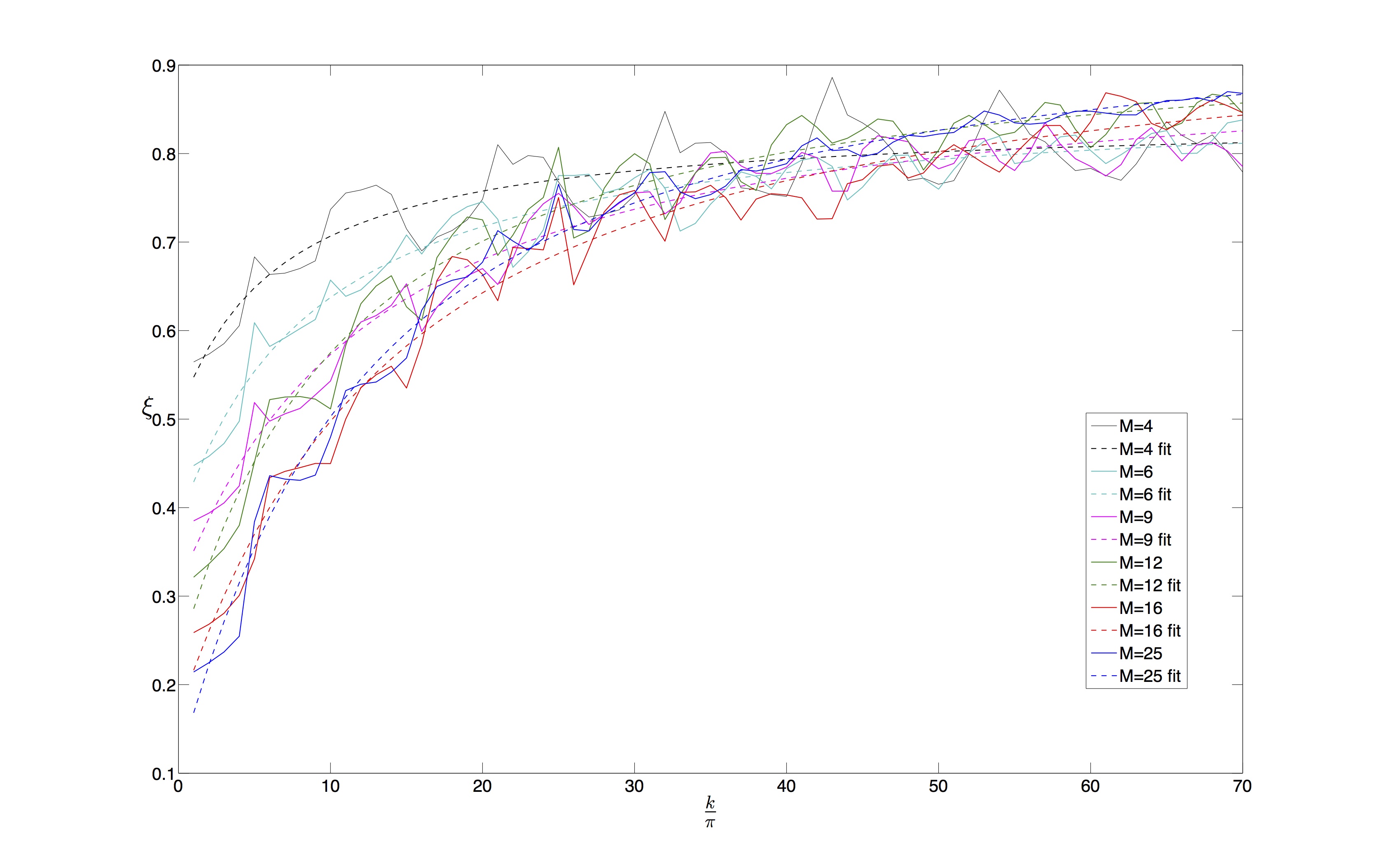}
\caption{\label{fig3}Results for $M=4,6,9,12,16,25$ (omitting the `anomalous' case
$M=20$). For each $M$ we plot the curve $\xi(k)$ obtained from simulations
(solid line) and the best fit to the function (\ref{fit1}) (dashed line).}
\end{center}
\end{figure}

Let us then examine the results of best-fits to the function (\ref{fit1}) for
$M=4,6,9,12,16,25$ (omitting $M=20$). The results are shown in Figure \ref{fig3}. For
each $M$ we display the curve $\xi(k)$ obtained from the simulations together
with the best-fit curve. We find good fits to the function (\ref{fit1}) on the
whole interval $(\pi,70\pi)$, with oscillations around the curve.

As is plain from Figure \ref{fig3}, the simulated functions $\xi(k)$ have an
oscillatory structure around a smooth curve that is well-approximated by
(\ref{fit1}). We may then study how the best-fit parameters $c_{1},c_{2},c_{3}$ vary with $M$.

The numerical values obtained for $c_{1},c_{2},c_{3}$ as $M$ varies are listed
in Table \ref{table1}. As $M$ increases, $c_{1}$ and $c_{2}$ decrease more or less
monotonically while $c_{3}$ steadily approaches $1$.

\begin{table}
\begin{center}
\begin{tabular}{ | c | c | c | c | }
\hline
$M$ & $c_1$ &  $c_2$ & $c_3$  \\                    
\hline
$4$ & $0.46$ &  $2.85$ & $0.84$  \\        
$6$ & $0.24$ &  $1.91$ & $0.86$  \\    
$9$ & $0.14$ &  $1.44$ & $0.92$  \\    
$12$ & $0.14$ &  $1.14$ & $0.95$  \\    
$16$ & $0.10$ &  $0.97$ & $0.97$  \\    
$25$ & $0.11$ &  $0.83$ & $0.99$  \\
\hline    
\end{tabular}
\caption{\label{table1}Results for the best-fit parameters $c_{1}$, $c_{2}$ and $c_{3}$ for
varying $M$ (with the fixed time interval $(t_{i},t_{f})=(10^{-4},10^{-2})$).}
\end{center}
\end{table}

Given the best-fit values $c_{1},c_{2},c_{3}$ for varying $M$, we now examine
how these values may be fit to simple functions $c_{1}=c_{1}(M)$, $c_{2}
=c_{2}(M)$, $c_{3}=c_{3}(M)$.

\begin{figure}
\begin{center}
\includegraphics[width=0.9\textwidth]{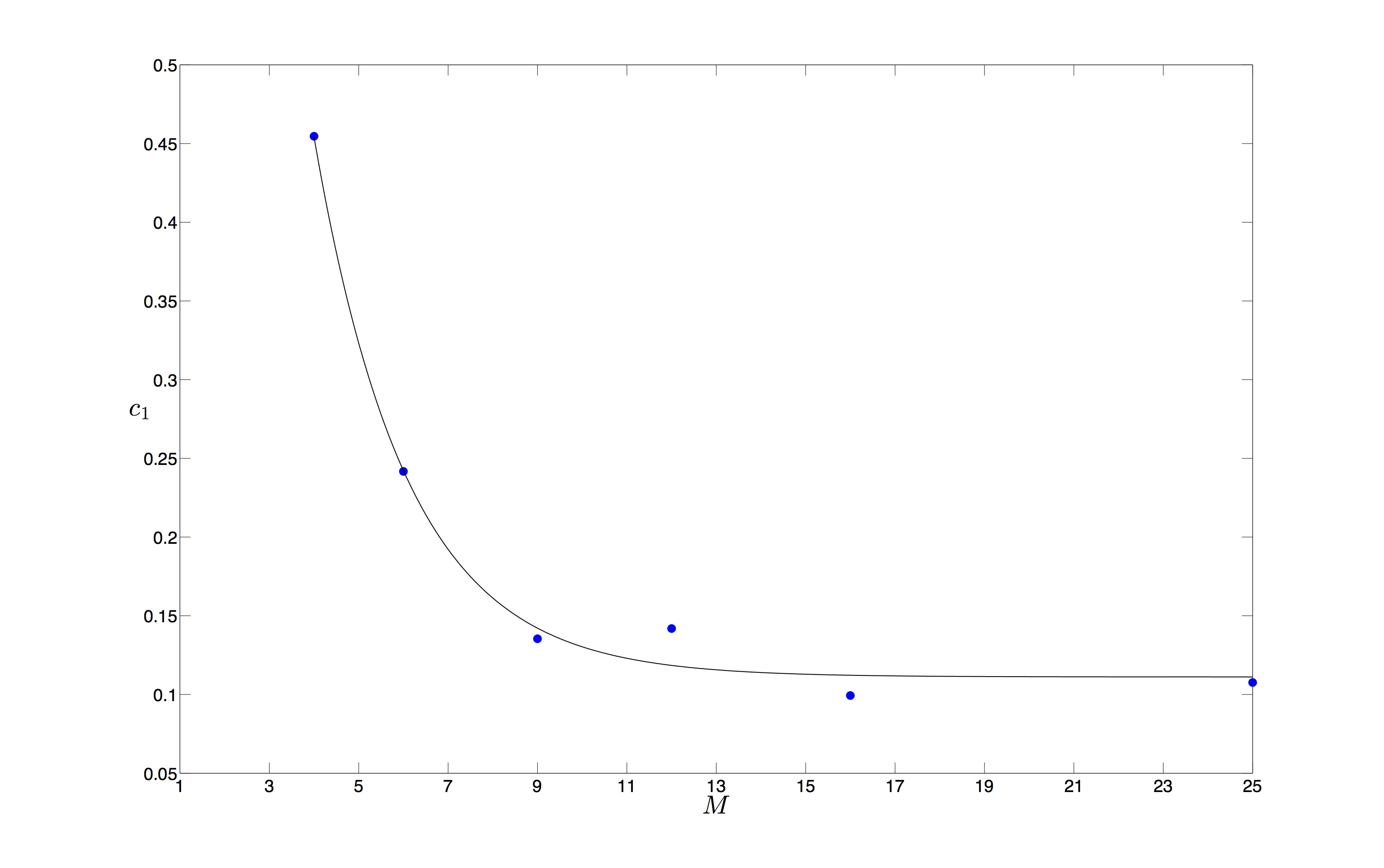}
\caption{\label{fig4}Plot of $c_{1}$ as a function of $M$, with a best-fit curve
(\ref{c1fit0}).}
\end{center}
\end{figure}

For $c_{1}$ as a function of $M$ we find a good fit to the curve
\begin{equation}
c_{1}=0.11+2.35e^{-0.48M}\ . \label{c1fit0}
\end{equation}
This is shown in Figure \ref{fig4}, which includes data points at $M=4,6,9,12,16,25$
(omitting the `anomalous' case $M=20$).

\begin{figure}
\begin{center}
\includegraphics[width=0.9\textwidth]{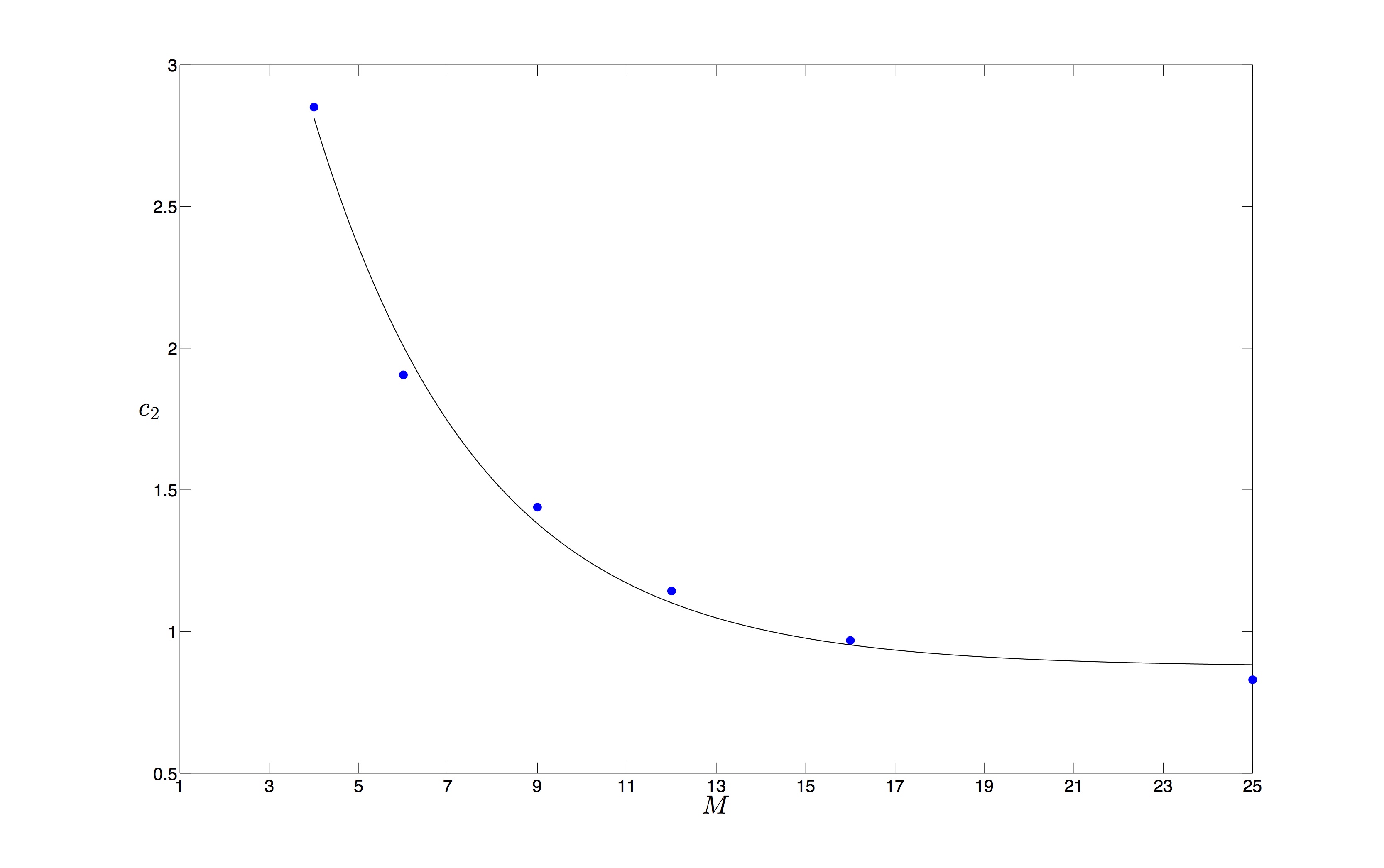}
\caption{\label{fig5}Plot of $c_{2}$ as a function of $M$, with a best-fit curve (\ref{c2fit}).}
\end{center}
\end{figure}

For $c_{2}$ as a function of $M$ we find a good fit to
\begin{equation}
c_{2}=0.88+5.67e^{-0.27M}\ . \label{c2fit}
\end{equation}
This is shown in Figure \ref{fig5}, again for $M=4,6,9,12,16,25$.

\begin{figure}
\begin{center}
\includegraphics[width=0.9\textwidth]{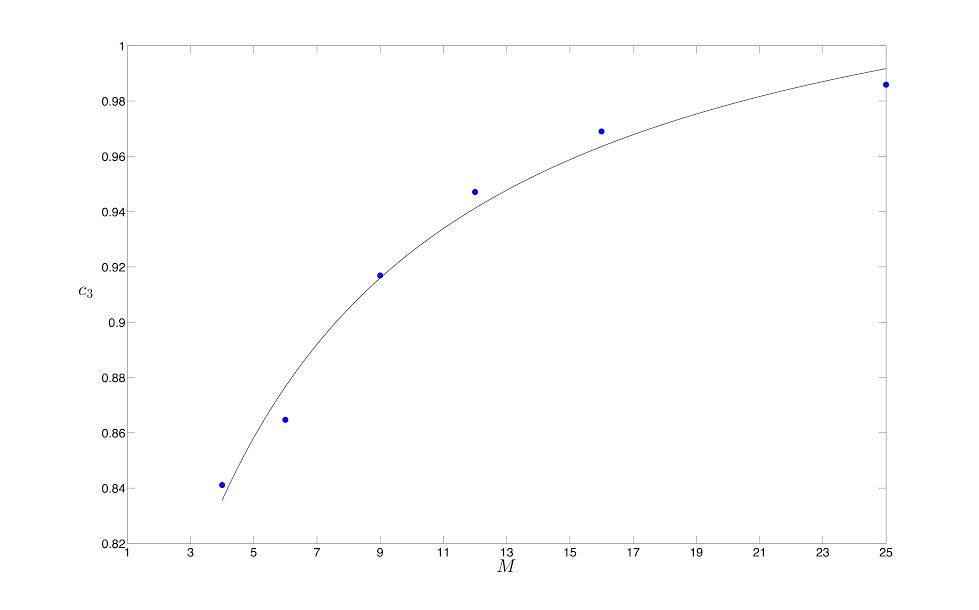}
\caption{\label{fig6}Plot of $c_{3}$ as a function of $M$, with a best-fit curve
(\ref{c3fit}).}
\end{center}
\end{figure}

Finally, for $c_{3}$ as a function of $M$ we find a good fit to
\begin{equation}
c_{3}=\tan^{-1}(0.53M+2.36)-\frac{\pi}{2}+1.06\ . \label{c3fit}
\end{equation}
This is shown in Figure \ref{fig6}  (for $M=4,6,9,12,16,25$).

As we have noted, the parameter $c_{3}$ is the limiting value (or `residue')
of $\xi$ as $k\rightarrow\infty$. According to (\ref{c3fit}), this parameter
in turn has a limiting value $c_{3}\rightarrow c\simeq1$ as $M\rightarrow
\infty$. (The difference of the fitted value $c=1.06$ from $1$ does not seem
sufficiently large to be considered significant, given the accuracy of our
simulations and of our fits.) It then appears that, as best as we are able to
determine, in the limiting regime where the wavenumber and the number of modes
are both large we will find that $\xi\simeq1$ -- that is, equilibrium will be
obtained to good accuracy, as one expects in a regime with both short
wavelengths and large numbers of modes.

One might ask why $c_{3}$ is not equal to $1$ for small values of $M$. The
observed behaviour of $c_{3}$ is in fact consistent with what is already known
about relaxation. In ref. \cite{LongRel} it was shown that, for a standard
two-dimensional harmonic oscillator (which corresponds mathematically to the
short-wavelength or large $k$ limit for our field mode on expanding space), if
the number $M$ of energy states in the superposition is small then while
relaxation takes place to a good approximation it is unlikely to take place
completely. This is because the trajectories are unlikely to fully explore the
configuration space, resulting in a small `residue' in the coarse-grained
$H$-function (indicating a small deviation from equilibrium) even in the
long-time limit. Whether or not there is a residue depends on the relative
phases in the initial superposition. If these are chosen randomly, then
long-time simulations indicate that a nonequilibrium residue is likely to
exist for small $M$ and unlikely to exist for large $M$ (see ref.
\cite{LongRel} for details). For small $M$, then, we may expect a similar
nonequilibrium `residue' in the width of the distribution at large $k$. Thus
it may be expected that $c_{3}$ will be slightly less than $1$ (noting that
our results are obtained by averaging over mixed states with randomly-chosen
initial phases) and that $c_{3}$ will become closer to $1$ for larger $M$ --
as indeed is observed in our results (Figure \ref{fig6}).

\subsection{Varying time interval and fixed number of modes}

\begin{figure}
\begin{center}
\includegraphics[width=0.9\textwidth]{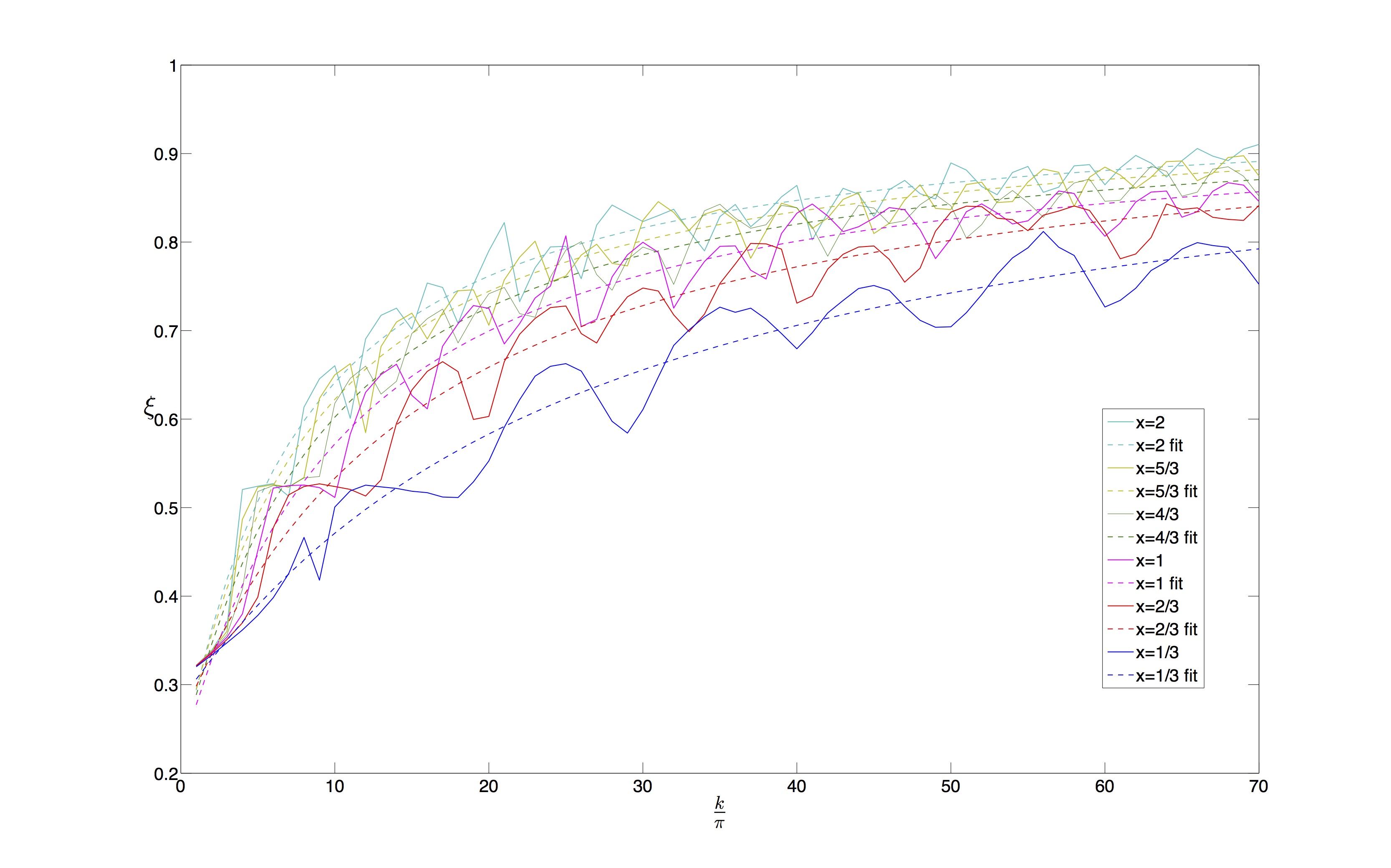}
\caption{\label{fig7}Results for $\xi(k)$ with varying final time $t_{f}=0.01x$ (where
$x=1/3,\ 2/3,\ 1,\ 4/3,\ 5/3,\ 2$) and a fixed number $M=12$ of modes. The
initial time $t_{i}=10^{-4}$ is kept fixed. Also displayed are best fits to
the function (\ref{fit1}).}
\end{center}
\end{figure}

We have also performed simulations for $\xi(k)$ with a varying final time
$t_{f}$ (the initial time $t_{i}=10^{-4}$ is kept fixed) and with a fixed
number $M=12$ of modes. In Figure \ref{fig7}  we display our results for $t_{f}=0.01x$,
where $x=1/3,\ 2/3,\ 1,\ 4/3,\ 5/3,\ 2$, together with best-fits to the
function (\ref{fit1}).

There is again a good fit to the function (\ref{fit1}). We see that, as
$t_{f}$ increases, $\xi(k)$ increases overall. This is expected, since a
longer time interval will in general yield more relaxation for fixed $M$. (As
before, to a first approximation we may ignore the oscillations. We note,
however, that as $t_{f}$ increases the period of the oscillations decreases.)

The values of the best-fit parameters $c_{1}$, $c_{2}$ and $c_{3}$ for varying
$t_{f}$ are shown in Table \ref{table2}. We see that $c_{1}$ and $c_{2}$ vary
significantly with $t_{f}$, while $c_{3}$ is essentially constant.

\begin{table}
\begin{center}
\begin{tabular}{ | c | c | c | c | }
\hline
$t_f$ & $c_1$ &  $c_2$ & $c_3$  \\                    
\hline
$0.01/3$ & $0.06$ &  $1.21$ & $0.97$  \\        
$0.02/3$ & $0.10$ &  $1.18$ & $0.96$  \\    
$0.01$ & $0.14$ &  $1.14$ & $0.95$  \\    
$0.04/3$ & $0.16$ &  $1.12$ & $0.95$  \\    
$0.05/3$ & $0.18$ &  $1.11$ & $0.96$  \\    
$0.02$ & $0.20$ &  $1.08$ & $0.96$  \\
\hline    
\end{tabular}
\caption{\label{table2}Best-fit parameters $c_{1}$, $c_{2}$, $c_{3}$ for varying $t_{f}$
(with fixed $t_{i}=10^{-4}$, $M=12$).}
\end{center}
\end{table}

\begin{figure}
\begin{center}
\includegraphics[width=0.9\textwidth]{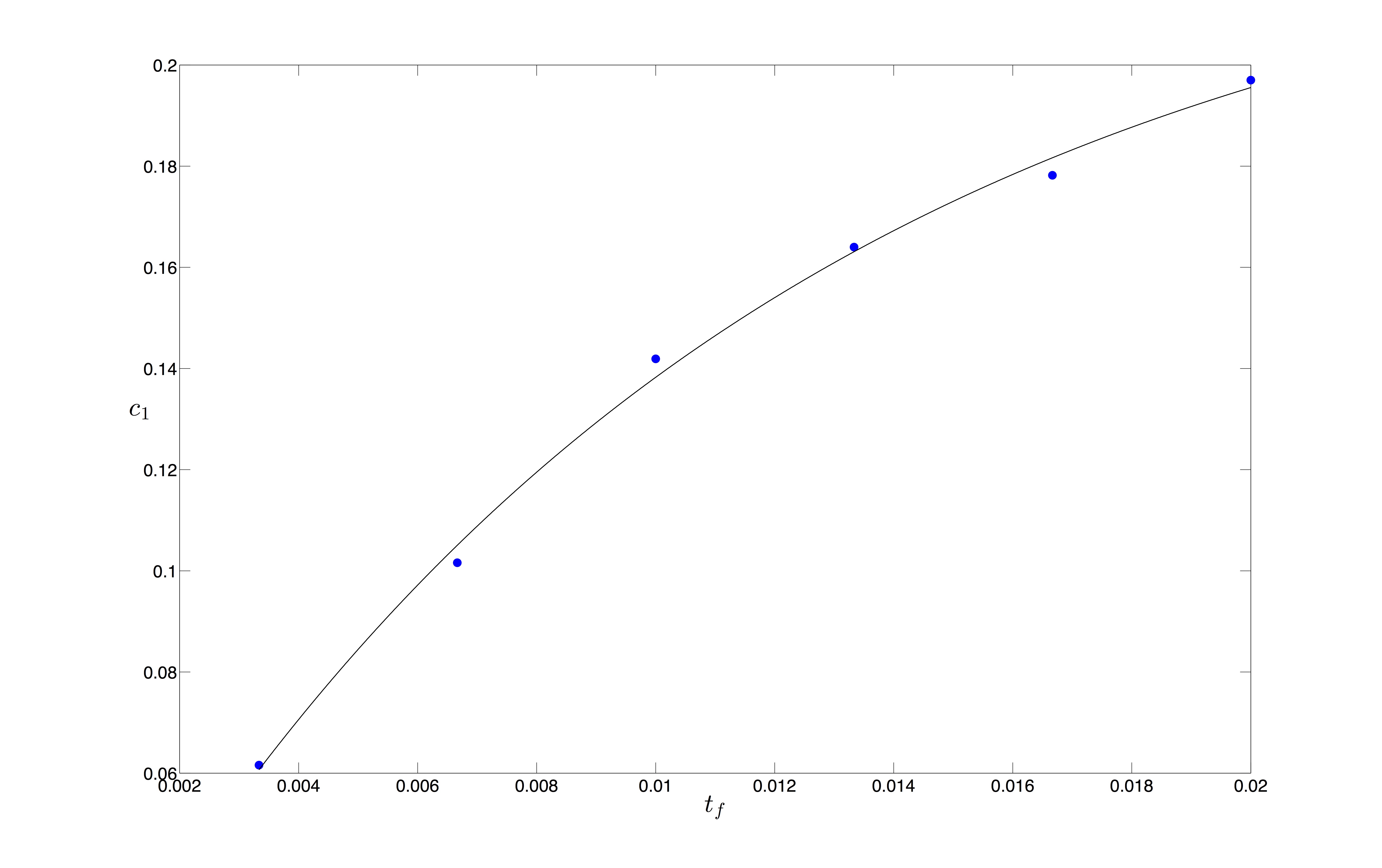}
\caption{\label{fig8}Plot of $c_{1}$ as a function of $t_{f}$, together with a best-fit to
the curve (\ref{c1fit'}).}
\end{center}
\end{figure}

In Figure \ref{fig8} we plot $c_{1}$ as a function of $t_{f}$, together with a
best-fit to the curve
\begin{equation}
c_{1}=0.24-0.24e^{-87.16t_{f}}\ . \label{c1fit'}
\end{equation}
Note that for $t_{f}=t_{i}$ the curve (\ref{c1fit'}) yields $c_{1}\simeq0$.
This is consistent with the fact that the `initial' ratio $\xi(k)$ -- defined
as above but with $t_{f}=t_{i}$ -- is independent of $k$. Our initial
distributions $\rho(q_{1},q_{2},t_{i})$ and $|\psi(q_{1},q_{2},t_{i})|^{2}$
all have variances that are proportional to $1/k$ and so the dependence on $k$
cancels out in the initial ratio $\xi(k)$ (which for $M=12$ is found to be
$\simeq0.3$ for all $k$).

\begin{figure}
\begin{center}
\includegraphics[width=0.9\textwidth]{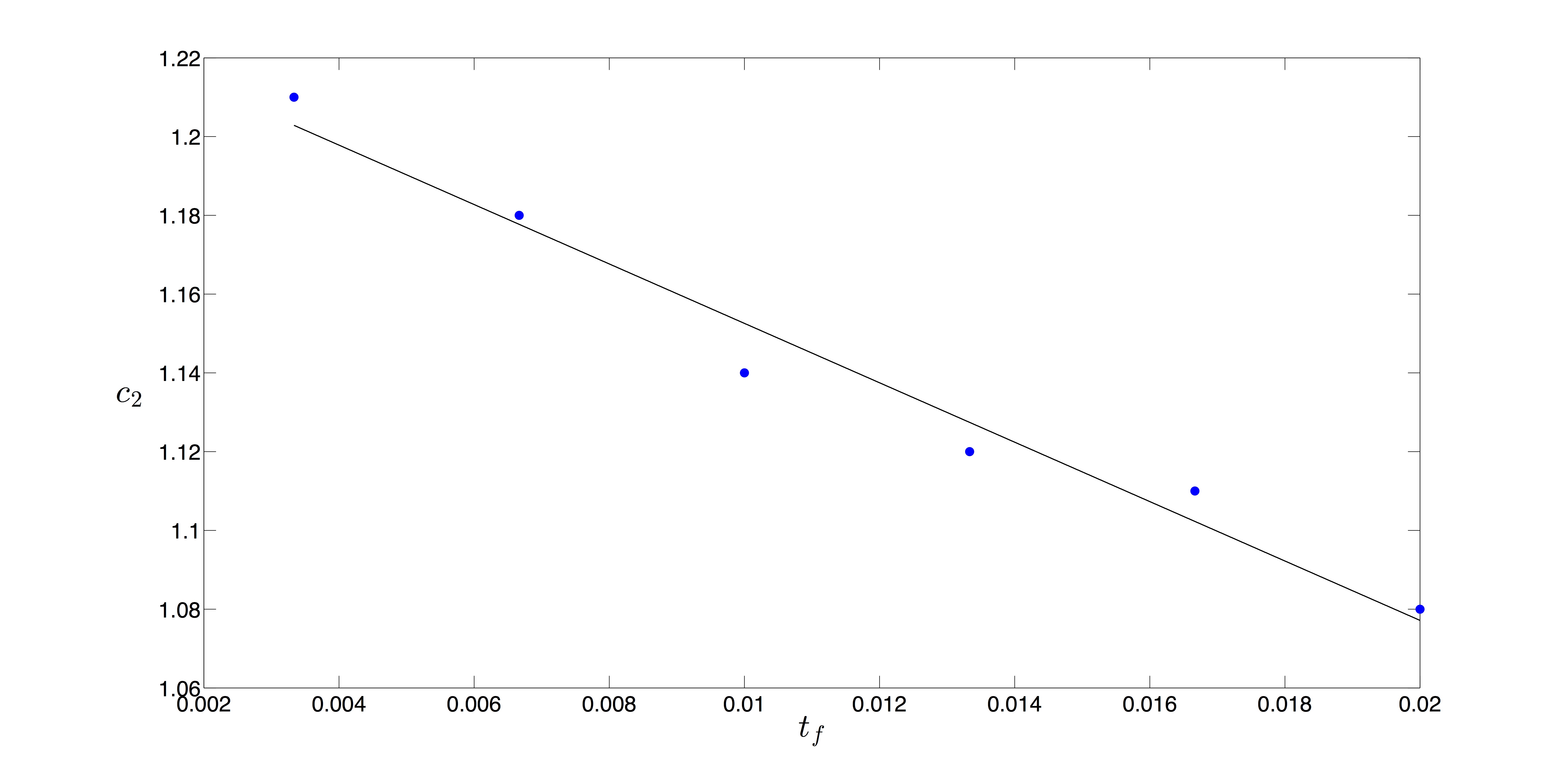}
\caption{\label{fig9}Plot of $c_{2}$ as a function of $t_{f}$, together with a best-fit to
the straight line (\ref{c2fit'}).}
\end{center}
\end{figure}

In Figure \ref{fig9}  we plot $c_{2}$ as a function of $t_{f}$, together with a
best-fit to the straight line
\begin{equation}
c_{2}=1.23-7.54t_{f}\ . \label{c2fit'}
\end{equation}

A fuller exploration of the range of parameters $t_{f}$, $M$ must be left for
future work, perhaps with greater computational resources. In the regime
studied here, $c_{1}$ and $c_{2}$ depend on both $M$ and $t_{f}$ while $c_{3}$
depends (essentially) only on $M$. We have found best-fit functions
$c_{1}=c_{1}(M)$, $c_{2}=c_{2}(M)$, $c_{3}=c_{3}(M)$ (equations (\ref{c1fit0}
), (\ref{c2fit}) and (\ref{c3fit})) for fixed $t_{f}=10^{-2}$ and $c_{1}
=c_{1}(t_{f})$, $c_{2}=c_{2}(t_{f})$ (equations (\ref{c1fit'}) and
(\ref{c2fit'})) for fixed $M=12$. It is evident that we really have a model
with two parameters $M$ and $t_{f}$ -- assuming a fixed initial time $t_{i}$
and a given initial nonequilibrium distribution (\ref{rhoinitial}). However,
further simulations that explore the $M-t_{f}$ plane are required to find the
best-fit dependence $c_{1}=c_{1}(M,t_{f})$ and $c_{2}=c_{2}(M,t_{f})$, as well
as to confirm if $c_{3}$ is essentially independent of $t_{f}$ while varying
with $M$. Given the three parameters $c_{1},c_{2},c_{3}$ as functions of
$M,t_{f}$, we would then have an explicit two-parameter model for the power deficit.

\subsection{Oscillations in the power deficit}

Our simulated deficit functions $\xi(k)$ show oscillations. As a first
approximation we have ignored these and found fits to the inverse-tangent
function (\ref{fit1}). Here we attempt to find fits that capture the
oscillations as well.

The oscillations in $\xi(k)$ may be related to the retarded time, which has an
oscillatory dependence on $k$ (see equations (\ref{epsilon})--(\ref{Thetaeval}
)). In effect, up to a final time $t_{f}$ our field system evolves like an
ordinary oscillator up to a final time $t_{\mathrm{ret}}(t_{f},k)$ that
depends on $k$. As $t_{\mathrm{ret}}(t_{f},k)$ rises or falls with varying
$k$, we broadly expect a larger or smaller degree of relaxation respectively.
Since the ordinary oscillator shows an exponential decay of the coarse-grained
$H$-function $\bar{H}(t)$ with time $t$, and since $\xi$ approaches $1$ as the
system relaxes, it is heuristically natural to attempt a fit of the form
\begin{equation}
\xi(k)=a-b\exp(-ct_{\mathrm{ret}}(t_{f},k))\ . \label{oscfit1}
\end{equation}

\begin{figure}
\begin{center}
\includegraphics[width=0.9\textwidth]{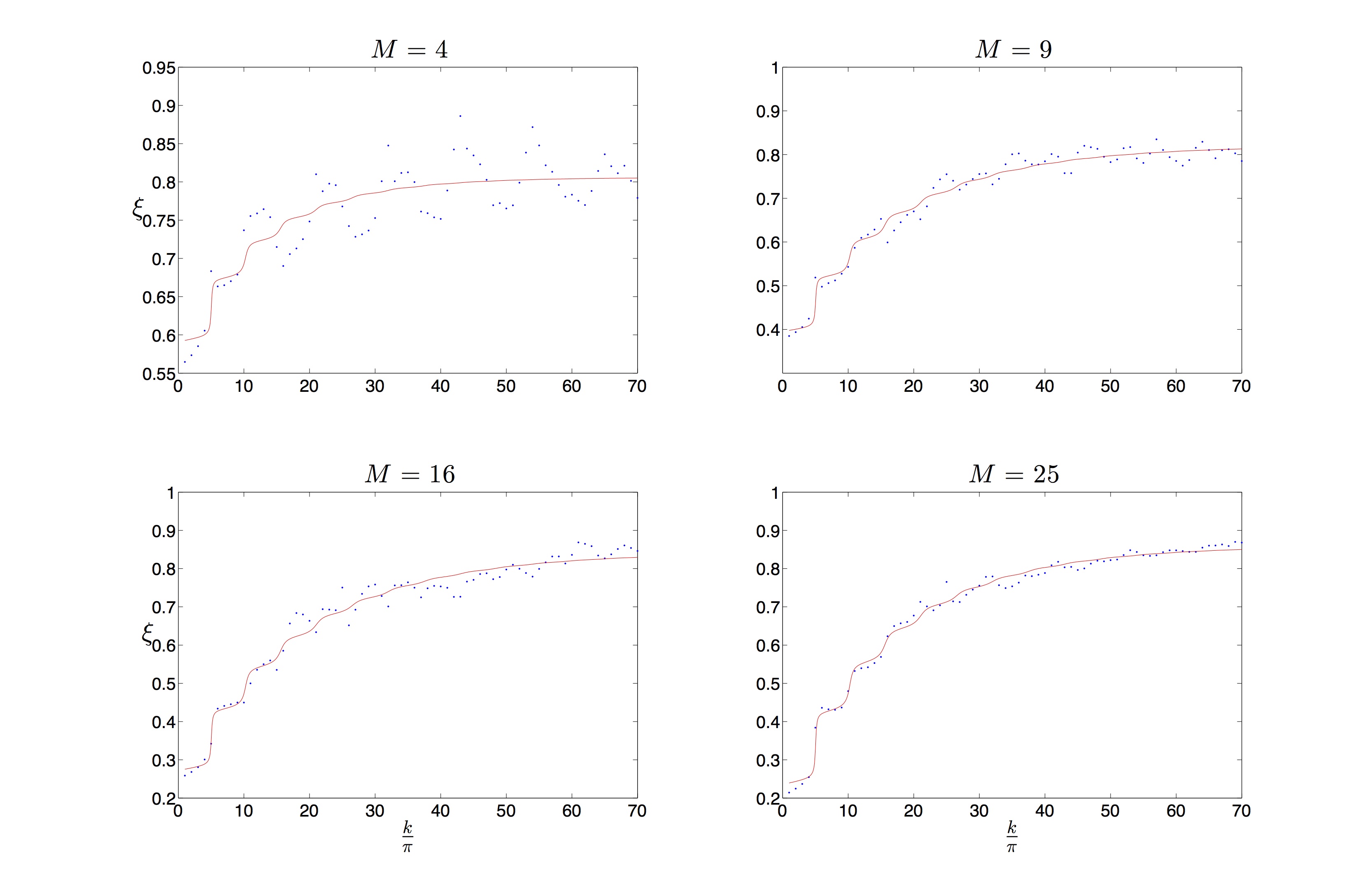}
\caption{\label{fig10}Fitting the oscillations in the power deficit with the exponential
function (\ref{oscfit1}) of the retarded time $t_{\mathrm{ret}}(t_{f},k)$
(with varying $M$ and fixed $t_{f}=10^{-2}$).}
\end{center}
\end{figure}

Best fits to the function (\ref{oscfit1}) have been performed for varying
$M=4,6,9,12,16,20,25$ (with fixed $t_{f}=10^{-2}$). Illustrative results for
the cases $M=4,9,16,25$ are shown in Figure \ref{fig10}.

We see that the fit (\ref{oscfit1}) captures the overall shape of the curve
just as well as the inverse-tangent fit (\ref{fit1}), while in addition
capturing some features of the oscillations in particular at low $k$. The fit
to the oscillations is better for larger $M$. However, the fit to the
oscillations is poor at very low $M$ (where as we have noted very low $M$ is
probably most realistic for the pre-inflationary phase).

As $M$ increases from $M=4$ to $M=25$ (with fixed $t_{f}=10^{-2}$) the
best-fit values of $a$, $c$ are roughly constant while $b$ increases
monotonically (the case $M=20$ again being an exception, see Table \ref{table3}).

\begin{table}
\begin{center}
\begin{tabular}{ | c | c | c | c | }
\hline
$M$ & $a$ &  $b$ & $c$  \\                    
\hline
$4$ & $0.81$ &  $0.21$ & $0.14$  \\        
$6$ & $0.80$ &  $0.32$ & $0.12$  \\    
$9$ & $0.82$ &  $0.42$ & $0.10$  \\    
$12$ & $0.85$ &  $0.50$ & $0.11$  \\    
$16$ & $0.84$ &  $0.57$ & $0.10$  \\    
$20$ & $0.92$ &  $0.64$ & $0.09$  \\
$25$ & $0.86$ &  $0.62$ & $0.11$  \\
\hline    
\end{tabular}
\caption{\label{table3}Best-fit parameters $a$, $b$, $c$ for the curve (\ref{oscfit1})
(varying $M$ and fixed $t_{f}=10^{-2}$).}
\end{center}
\end{table}

\begin{figure}
\begin{center}
\includegraphics[width=0.9\textwidth]{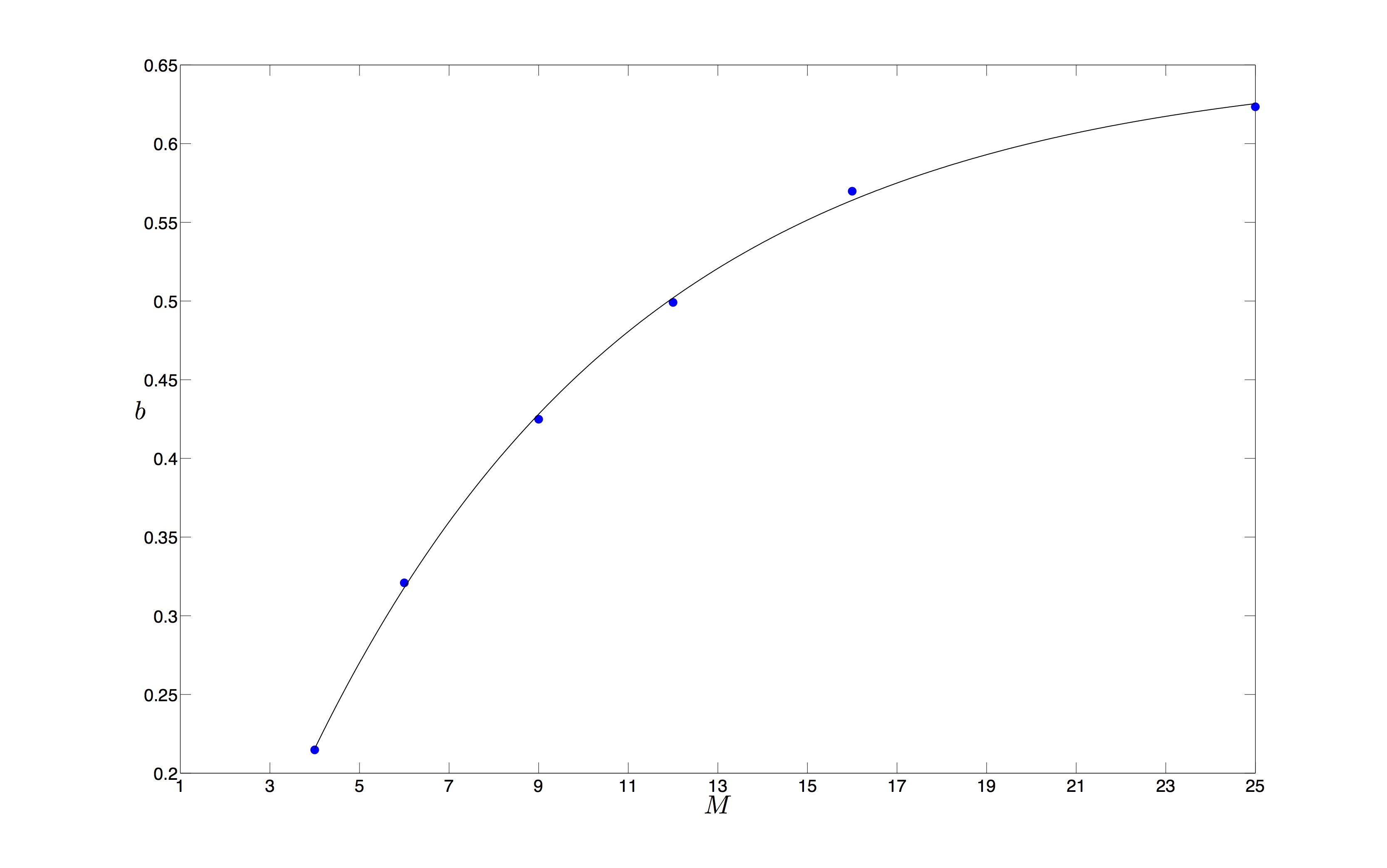}
\caption{\label{fig11}The parameter $b$ as a function of $M$, with a best-fit curve
(\ref{b(M)}).}
\end{center}
\end{figure}

In Figure \ref{fig11}  we plot $b$ as a function of $M$ with a best-fit curve
\begin{equation}
b=0.65-0.74\exp(-0.13M) \label{b(M)}
\end{equation}
(omitting the `anomalous' spike at $M=20$).

Best fits to the curve (\ref{oscfit1}) have also been evaluated for varying
$t_{f}=0.01x$ where $x=1/3,\ 2/3,\ 1,\ 4/3,\ 5/3,\ 2$ (with fixed $M=12$). The
best-fit values of $a$, $b$, $c$ are very nearly constant over this range of
$t_{f}$, with $a$ and $b$ steadily increasing by a small fraction while $c$
decreases slightly (see Table \ref{table4}). Thus to a first approximation $\xi$
depends on $t_{f}$ only via the known function $t_{\mathrm{ret}}(t_{f},k)$.
(Note that the expression for $t_{\mathrm{ret}}(t_{f},k)$, as defined by
equations (\ref{epsilon})--(\ref{Thetaeval}), is independent of $M$.)%

\begin{table}
\begin{center}
\begin{tabular}{ | c | c | c | c | }
\hline
$t_f$ & $a$ &  $b$ & $c$  \\                    
\hline
$0.01/3$ & $0.83$ &  $0.47$ & $0.12$  \\        
$0.02/3$ & $0.84$ &  $0.49$ & $0.11$  \\    
$0.01$ & $0.85$ &  $0.50$ & $0.11$  \\    
$0.04/3$ & $0.86$ &  $0.50$ & $0.11$  \\    
$0.05/3$ & $0.87$ &  $0.52$ & $0.10$  \\    
$0.02$ & $0.88$ &  $0.52$ & $0.10$  \\
\hline    
\end{tabular}
\caption{\label{table4}Best-fit parameters $a$, $b$, $c$ for the curve (\ref{oscfit1})
(varying $t_{f}$ and fixed $M=12$).}
\end{center}
\end{table}

We again have a two-parameter model of the power deficit in terms of
parameters $M$, $t_{f}$ (again for a fixed initial time $t_{i}$ and a given
initial nonequilibrium distribution (\ref{rhoinitial})).

\begin{figure}
\begin{center}
\includegraphics[width=0.9\textwidth]{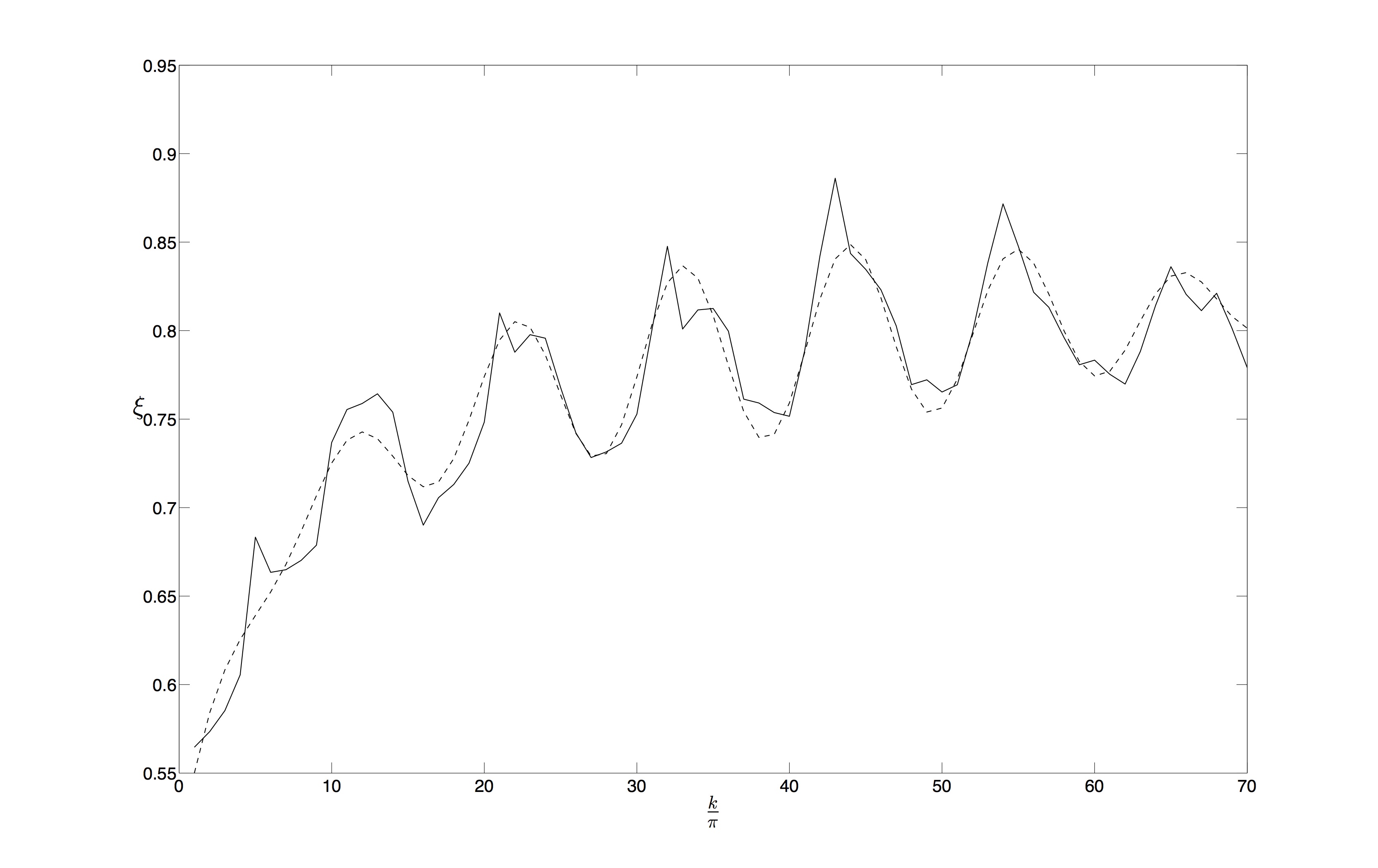}
\caption{\label{fig12}An alternative best-fit curve (\ref{oscfit2}) for the case $M=4$
(with coefficients $c_{1}=0.46$, $c_{2}=2.87$, $c_{3}=0.84$, $c_{4}=0.05$,
$c_{5}=0.57$, $c_{6}=0.04$).}
\end{center}
\end{figure}

The fit (\ref{oscfit1}) provides an approximate account of the oscillations in
$\xi(k)$ for the low-$k$ region. However it does not at all capture the
oscillations in $\xi(k)$ for higher $k$, which are especially large for very
low $M$. We have tried an alternative fit of the form
\begin{equation}
\xi(k)=\tan^{-1}(c_{1}k/\pi+c_{2})-\pi/2+c_{3}+c_{4}\cos(c_{5}k/\pi)\sin
(c_{6}k/\pi)\ , \label{oscfit2}
\end{equation}
which simply adds an oscillatory function to the inverse tangent
(\ref{fit1}).\footnote{This suffices in the studied $k$-region, though the
sine function might be replaced by a Gaussian so as to damp away the
oscillations at $k>70\pi$.} We find a fairly good fit for $M=4$, as shown in
Figure \ref{fig12}, but not for $M=6$ or $M=9$. For the latter cases Fourier analysis
shows the presence of additional frequencies. (As we have noted, by eye one
sees from Figure \ref{fig2} that the oscillations in $\xi(k)$ can be rather erratic
even if they appear regular for $M=4$.)

A full characterisation of the oscillations is left for future work. However,
we may make some general comments. First, oscillations in the large-scale
(low-$k$) primordial spectrum appear to be a generic prediction of our model
(although further study is required to fully parameterise their features).
Second, oscillations in the primordial spectrum also seem to be a generic
prediction of models with trans-Planckian corrections to quantum field theory
\cite{BM13}. How these (generally differing) predictions may be compared and
distinguished is left for future work. Finally, the overall success of the fit
(\ref{oscfit1}) -- which performs at least as well as the inverse-tangent fit
(\ref{fit1}) -- suggests that the power deficit will approximately follow this
generic form for arbitrary cosmological expansions and not just for a
radiation-dominated expansion (where different functions of time for the scale
factor $a=a(t)$ will imply different functions for the retarded time
$t_{\mathrm{ret}}(t,k)$).

\section{Phase relaxation}

So far we have studied the nonequilibrium deficit function $\xi(k)$, which
measures the deviation of the width of the primordial distribution from the
equilibrium value. The observed power deficit at $l\lesssim40$ might be caused
by a dip in $\xi(k)$ at long wavelengths (small $k$).

Also of interest is the distribution of the \textit{phase} $\theta
_{\mathbf{k}}=\tan^{-1}(q_{\mathbf{k}2}/q_{\mathbf{k}1})$ of the Fourier
component of the primordial field. The observed anisotropy in the CMB at
$l\lesssim10$ might be caused by nonequilibrium phases at long wavelengths.

As we have noted, statistical isotropy for the CMB implies the standard
relations (\ref{iso}). (These might be satisfied, for example, by a Gaussian
field with uncorrelated phases.) Isotropy therefore requires that
$\left\langle a_{l^{\prime}m^{\prime}}^{\ast}a_{lm}\right\rangle =0$ for all
$l,m\neq l^{\prime},m^{\prime}$. However, data from the Planck satellite show
evidence for `phase correlations' at low $l$, in the sense that $\left\langle
a_{l^{\prime}m^{\prime}}^{\ast}a_{lm}\right\rangle \neq0$ for some $l,m\neq
l^{\prime},m^{\prime}$ at low values of $l,l^{\prime}$ (in the region
$l,l^{\prime}\lesssim10$). The Planck team also report a seemingly anomalous
or unlikely mode alignment, as well as various other effects that indicate
statistical anisotropy \cite{PlanckXXIII}.

The reported phase correlations refer to the phases of the complex
coefficients $a_{lm}$, and not directly to the phases of the primordial
perturbations. From the linear expression (\ref{alm}) for $a_{lm}$ in terms of
$\mathcal{R}_{\mathbf{k}}$, it is clear that the phase of a given $a_{lm}$ is
in principle related to all of the primordial phases -- that is, to the phases
of all of the $\mathcal{R}_{\mathbf{k}}$'s or (equivalently) to the phases of
all of the $\phi_{\mathbf{k}}$'s. Writing $\phi_{\mathbf{k}}=\left\vert
\phi_{\mathbf{k}}\right\vert e^{i\theta_{\mathbf{k}}}$, it has been shown that
during inflation the phases $\theta_{\mathbf{k}}$ are static along the de
Broglie-Bohm trajectories for the inflaton perturbations, so that an initial
nonequilibrium distribution for the $\theta_{\mathbf{k}}$'s will remain
unchanged during the inflationary era \cite{AV10}. Thus, if there is an
anomalous distribution of primordial phases at the beginning of inflation,
this distribution will be preserved in time and transferred to cosmological
lengthscales (as occurs with the power deficit). The resulting anomalous
phases in the primordial curvature perturbations $\mathcal{R}_{\mathbf{k}}$
will then affect the observed phases of the coefficients $a_{lm}$ in the CMB.

It therefore seems important to study the relaxation of phases during our
radiation-dominated expansion, as a model of phase relaxation during a
possible pre-inflationary era, with a view to perhaps explaining the observed
anisotropy and associated phase correlations in the $a_{lm}$'s.

Thus we now study relaxation for the phases $\theta_{\mathbf{k}}$ associated
with our spectator scalar field on a radiation-dominated background. Consider
a mode of wave number $k$. We shall calculate the time evolution of the phase
marginal $\rho(\theta_{\mathbf{k}},t)$ -- that is, of the marginal probability
distribution for $\theta_{\mathbf{k}}$ (obtained by integrating over the
amplitude $\left\vert \phi_{\mathbf{k}}\right\vert $ in the total probability distribution).

For the initial (Gaussian) nonequilibrium distribution (\ref{rhoinitial}) the
phase marginal is uniform on the unit circle. Whereas for the initial wave
function (\ref{psiinitial}) the equilibrium phase marginal is non-uniform.
Thus we have a nonequilibrium phase marginal at the initial time
$t_{i}=10^{-4}$. We may then calculate the phase marginal at the final time
$t_{f}=10^{-2}$ for varying $M$ and $k$.

\begin{figure}
\begin{center}
\includegraphics[width=0.9\textwidth]{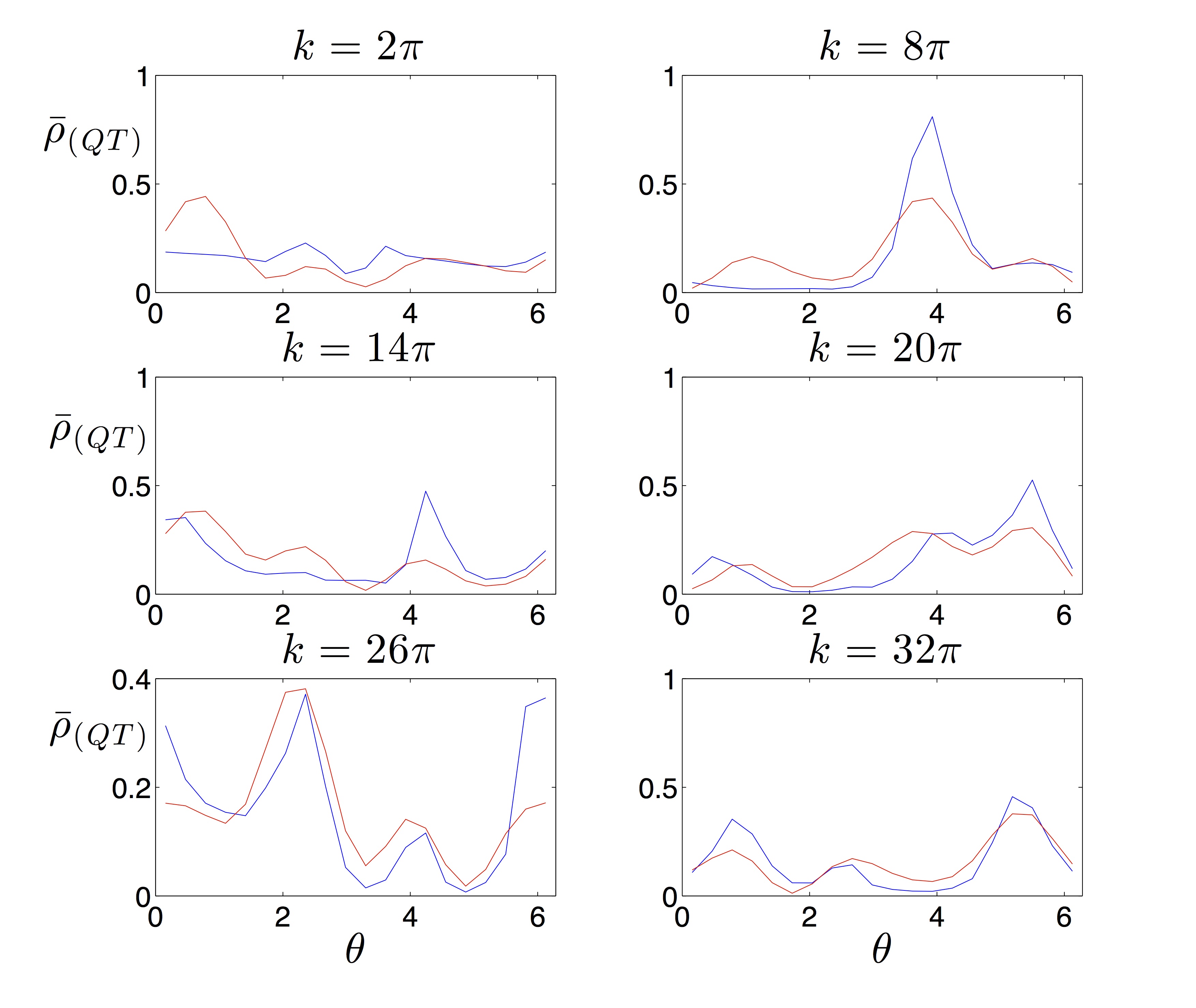}
\caption{\label{fig13}Phase relaxation for $M=4$ (final time $t_{f}=10^{-2}$ and one set of
initial phases in the wave function). The blue and red curves are the
nonequilibrium and equilibrium distributions respectively.}
\end{center}
\end{figure}

\begin{figure}
\begin{center}
\includegraphics[width=0.9\textwidth]{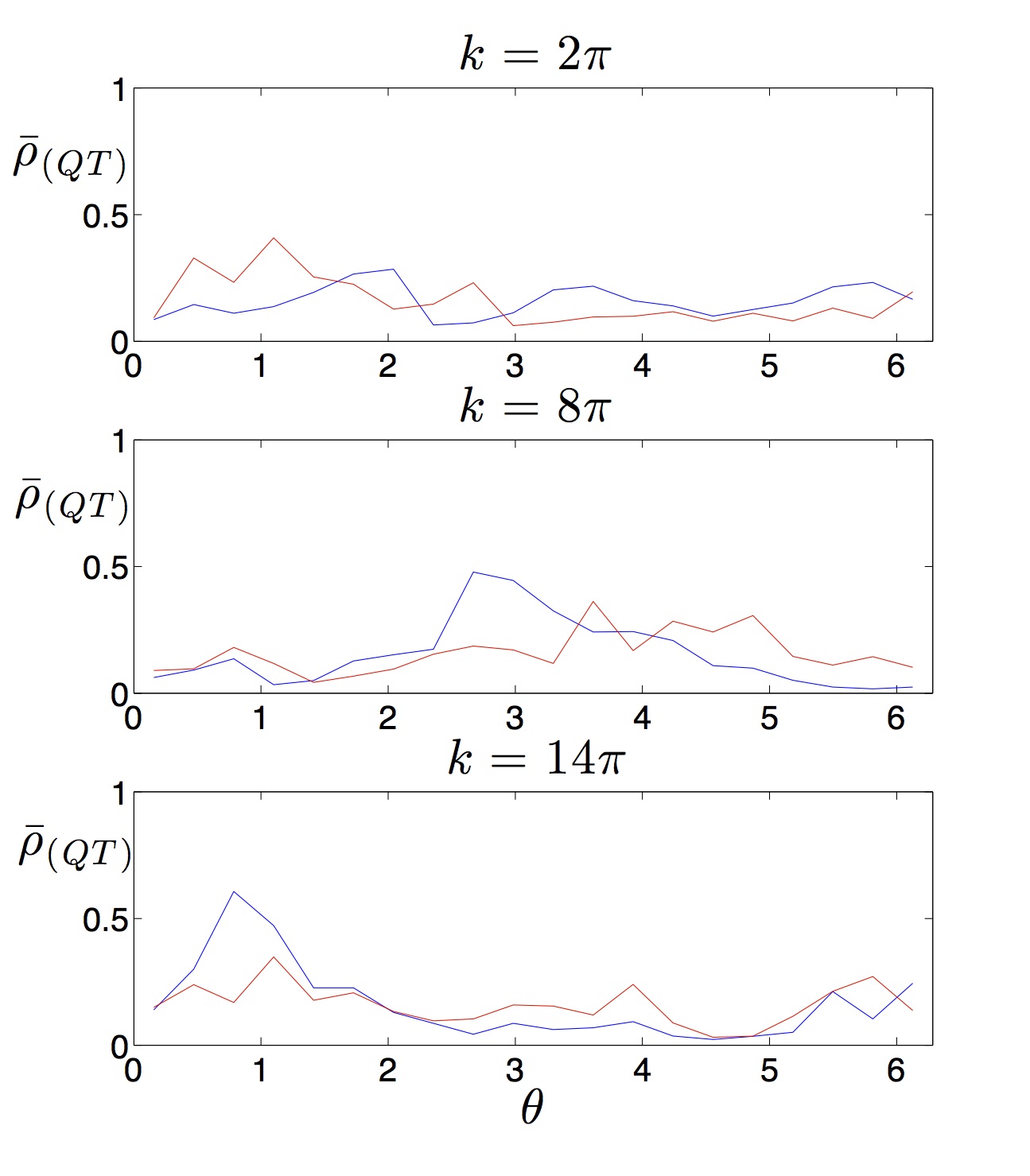}
\caption{\label{fig14}Phase relaxation for $M=25$ (final time $t_{f}=10^{-2}$ and one set
of initial phases in the wave function). The blue and red curves are the
nonequilibrium and equilibrium distributions respectively.}
\end{center}
\end{figure}

In Figures \ref{fig13} and \ref{fig14} we plot some illustrative results for the final
coarse-grained phase marginal $\bar{\rho}(\theta,t_{f})$ together with the final coarse-grained equilibrium marginal $\bar{\rho
}_{\mathrm{QT}}(\theta,t_{f})$ (omitting the label ${\mathbf{k}}$),\footnote{To calculate the coarse-grained
marginals, we first take 1001 equally-spaced angles at which we calculate the
fine-grained marginals by integrating the fine-grained joint distributions
radially. We then coarse-grain the results, yielding 20 coarse-grained values
for each distribution (where each value is an average over 51 fine-grained
values).} for $M=4$ and $M=25$ respectively, each with varying values of $k$.
(The set of initial phases in the wave function is fixed.) By eye one can
discern an approximate relaxation as $k$ increases.

We must quantify nonequilibrium in a way that is relevant to observations. For
amplitudes, observationally what matters is the width of the distribution and
so a calculation of $\xi(k)$ suffices. Whereas for phases, observationally
what matters is whether they are in quantum equilibrium or not. (For the
inflationary vacuum, equilibrium phases are uniformly distributed on
$(0,2\pi)$.) To quantify the deviation of the (coarse-grained) phase marginal
$\bar{\rho}(\theta,t)$ from the equilibrium marginal $\bar{\rho}_{\mathrm{QT}}(\theta,t)$ we may use the coarse-grained $H$-function
\begin{equation}
\bar{H}_{\theta}(t)=\int d\theta\ \bar{\rho}(\theta,t)\ln\left(  \bar{\rho
}(\theta,t)/\bar{\rho}_{\mathrm{QT}}(\theta,t)\right)  \ .
\end{equation}
Thus, while we use $\xi$ as a measure of nonequilibrium for amplitudes, we use
$\bar{H}_{\theta}$ as a measure of nonequilibrium for phases.

\begin{figure}
\begin{center}
\includegraphics[width=0.9\textwidth]{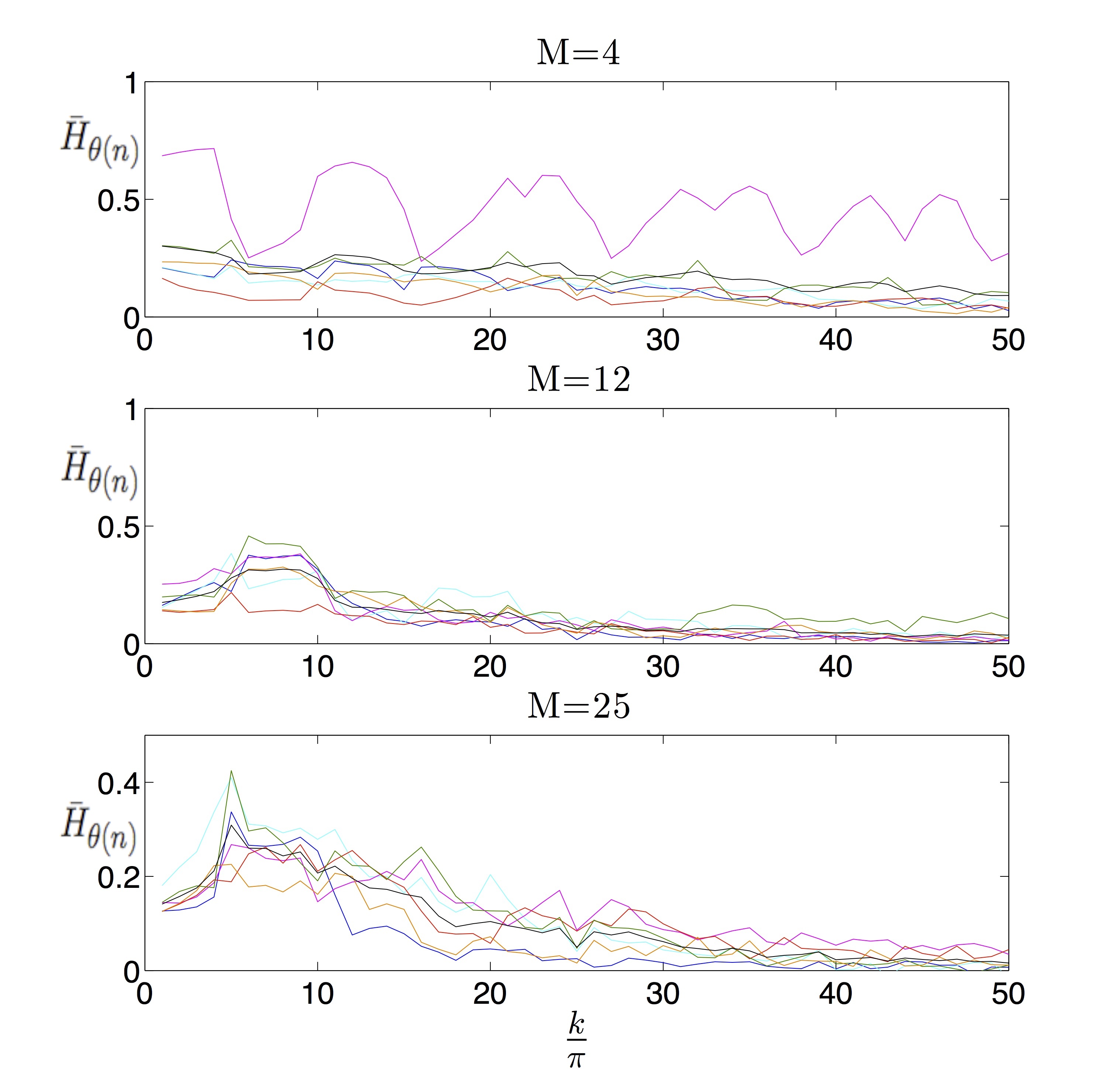}
\caption{\label{fig15}Coarse-grained $H$-functions for final phase marginals, for
$M=4,12,25$ (calculated with $t_{f}=10^{-2}$ and six sets of initial phases in
the wave function) and for $k$ in the range $(\pi,50\pi)$. The mean curve is
shown in black.}
\end{center}
\end{figure}

Of interest here is the process of relaxation towards quantum equilibrium,
$\bar{\rho}(\theta,t)\rightarrow\bar{\rho}_{\mathrm{QT}}(\theta,t)$, as
quantified by $\bar{H}_{\theta}(t)$. We consider the phase-marginal
$H$-function at the final time, $\bar{H}_{\theta}(t_{f})$, for varying $k$ and
$M$. Since $t_{f}=10^{-2}$ is fixed, for each $M$ we may regard $\bar
{H}_{\theta}$ as a function of $k$. The simulations are run for six sets of
initial phases in the wave function. For each $M$ we then obtain six separate
curves $\bar{H}_{\theta n}(k)$ (where as before the index $n$ labels the
initial wave function corresponding to the choice of initial phases). These
may be averaged to yield a mean curve, which we denote by $\bar{H}_{\theta
}(k)$. Some illustrative results are shown in Figure \ref{fig15} for $M=4$, $12$ and
$25$, each with $k$ in the range $(\pi,50\pi)$. Many of the curves show an
initial increase. Overall, however, there is a general decrease -- indicating
relaxation -- as $k$ increases. (Note that the initial increase is consistent,
since the ratio $\bar{\rho}(\theta,t)/\bar{\rho}_{\mathrm{QT}}(\theta,t)$ is
not conserved along trajectories for marginals and so the usual $H$-theorem
\cite{AV91a} cannot be derived for marginals.)

\section{Relaxation scales for amplitude and phase}

For the primordial perturbations we may define two critical $k$-scales,
$k_{\mathrm{amp}}$ and $k_{\mathrm{phase}}$, that characterise relaxation for
the amplitudes and phases respectively. We shall define $k_{\mathrm{amp}}$ as
a value of $k$ below which the amplitudes may roughly be said to have not
fully relaxed (so that there is a significant nonequilibrium power deficit).
Similarly, we define $k_{\mathrm{phase}}$ as a value of $k$ below which the
phases may roughly be said to have not fully relaxed (so that there is a
significant departure from quantum randomness). Precise definitions are given below.

The best-fit function (\ref{fit1}) for $\xi(k)$ has three parameters $c_{1}$,
$c_{2}$ and $c_{3}$. As we noted in Section 3.1, $\xi\rightarrow c_{3}<1$ as
$k\rightarrow\infty$, so that $c_{3}$ represents a nonequilibrium residue at
short wavelengths. We also found that $c_{3}\simeq1$ for large
$M$. But in a cosmology where the pre-inflationary era contains a small number
of modes, we may expect that $c_{3}$ is slightly less than $1$ (indicating a
slightly incomplete relaxation even at short wavelengths). Observationally
speaking, as already noted in Section 3.1, this would imply an overall renormalisation of the power spectrum,
in the sense that the value of $c_{3}$ would be absorbed into effective values
for other cosmological parameters (in particular the overall amplitude for the
power spectrum). Thus, a residual nonequilibrium $c_{3}<1$ in the power
spectrum for large $k$ would not by itself be noticeable. We might, however,
notice a dip in the function $\xi(k)$ for small values of $k$. Thus the
observable deficit in $\xi$ should be defined relative to the limiting value
$c_{3}$.

Let us then define the `renormalised' scale $k_{\mathrm{amp}}$ to be the value
of $k$ such that $\xi$ dips significantly below $c_{3}$. For example, we might
take
\begin{equation}
\xi(k_{\mathrm{amp}})=0.9c_{3}\;\mathrm{or}\;0.95c_{3}\ . \label{kamp}
\end{equation}
We shall consider both choices (the exact definition is of course a matter of
taste). We denote the resulting values by $k_{\mathrm{amp\ 10\%}}$ and
$k_{\mathrm{amp\ 5\%}}$ -- corresponding to $10\%$ and $5\%$ (primordial)
power deficits respectively.

We consider the $\xi$-curves that were obtained for fixed $t_{f}=10^{-2}$ and
varying $M$ (Section 3.1). For these cases the best-fit parameters $c_{1}$,
$c_{2}$, $c_{3}$ are listed in Table \ref{table1}. From (\ref{kamp}) we obtain the two
values $k_{\mathrm{amp\ 10\%}}$ and $k_{\mathrm{amp\ 5\%}}$ of the
characteristic $k$-scale, for each value of $M$ (again omitting the
`anomalous' case $M=20$). The results are displayed in Table \ref{table5}.
\begin{table}
\begin{center}
\begin{tabular}{ | c | c | c | }
\hline
$M$ & $k_{\mathrm{amp\ 10\%}}$ & $k_{\mathrm{amp\ 5\%}}$  \\                    
\hline
$4$ & $20\pi$ &  $46\pi$ \\        
$6$ & $40\pi$ &  $88\pi$  \\    
$9$ & $70\pi$ &  $150\pi$  \\    
$12$ & $66\pi$ &  $141\pi$  \\    
$16$ & $94\pi$ &  $198\pi$  \\    
$25$ & $86\pi$ &  $181\pi$  \\
\hline    
\end{tabular}
\caption{\label{table5}Results for the $k$-scales $k_{\mathrm{amp\ 10\%}}$ and
$k_{\mathrm{amp\ 5\%}}$ for varying $M$ (with the fixed time interval
$(t_{i},t_{f})=(10^{-4},10^{-2})$).}
\end{center}
\end{table}

The mean $k_{\mathrm{amp}}\equiv\frac{1}{2}(k_{\mathrm{amp\ 10\%}
}+k_{\mathrm{amp\ 5\%}})$ lies in the range
\begin{equation}
k_{\mathrm{amp}}\simeq33\pi-146\pi\ ,
\end{equation}
with the lowest values obtained for the lowest $M$ (specifically, $M=4$ and
$M=6$).

To define a phase relaxation scale $k_{\mathrm{phase}}$, we may use the mean
curve $\bar{H}_{\theta}(k)$ for the coarse-grained phase marginal. While the
curve $\bar{H}_{\theta}(k)$ shows an overall decrease with $k$, the dependence
is not exponential (see Figure \ref{fig15}). Even so, given an `initial' point $k_{i}$
we may define $k_{\mathrm{phase}}$ by
\begin{equation}
\bar{H}_{\theta}(k_{i}+k_{\mathrm{phase}})=(1/e)\bar{H}_{\theta}(k_{i})
\label{k_phase}
\end{equation}
(as one would if $\bar{H}_{\theta}$ were decaying exponentially on a
characteristic scale $k_{\mathrm{phase}}$).

Again focussing on the case of fixed $t_{f}=10^{-2}$ and varying $M$, we have
performed simulations for $M=4,12,16,20,25$ (each with six sets of initial
phases). For each $M$ we obtain an averaged curve $\bar{H}_{\theta}(k)$, from
which we may obtain a characteristic scale $k_{\mathrm{phase}}$ defined by
(\ref{k_phase}) (we take $k_{i}=\pi$). The results are: $k_{\mathrm{phase}
}=38\pi$, $25\pi$, $43\pi$, $32\pi$, $25\pi$ for $M=4,12,16,20,25$
respectively. Approximately, we find
\begin{equation}
k_{\mathrm{phase}}\simeq35\pi
\end{equation}
(varying with $M$ by about $30\%$).

Thus we find a ratio approximately in the range
\begin{equation}
\frac{k_{\mathrm{phase}}}{k_{\mathrm{amp}}}\simeq\allowbreak0.2-\allowbreak
1.1\ , \label{kratio}
\end{equation}
with the highest values obtained for the lowest $M$ (again $M=4$ and $M=6$).

Roughly speaking, primordial perturbations on a scale $k$ affect the CMB at a
multipole $l\simeq(2/H_{0})k$ (where $H_{0}$ is the Hubble parameter today)
\cite{LL00}. If we define analogous quantities $l_{\mathrm{amp}}$ and
$l_{\mathrm{phase}}$ as those values of $l$ below which we observe a power
deficit and phase anomalies respectively, then we should find a ratio
$l_{\mathrm{phase}}/l_{\mathrm{amp}}\simeq k_{\mathrm{phase}}/k_{\mathrm{amp}
}$ or
\begin{equation}
\frac{l_{\mathrm{phase}}}{l_{\mathrm{amp}}}\simeq0.2-\allowbreak1.1\ .
\label{lratio}
\end{equation}
The Planck data indicate values (roughly) of $l_{\mathrm{phase}}\simeq10$ and
$l_{\mathrm{amp}}\simeq40$, with a ratio
\begin{equation}
\frac{l_{\mathrm{phase}}}{l_{\mathrm{amp}}}\simeq0.25\ . \label{lratiodata}
\end{equation}
This seems reasonably consistent with our rough estimate (\ref{lratio}).

\section{Angular power deficit}

Let us study more precisely how the proposed deficit in the primordial power
spectrum could yield the reported deficit in the angular power spectrum at low
$l$.

At low $l$ the (square of the) transfer function takes the form \cite{LL00}
\begin{equation}
\mathcal{T}^{2}(k,l)=\pi H_{0}^{4}j_{l}^{2}(2k/H_{0})\ .
\end{equation}
As a first approximation let us assume that the quantum-theoretical primordial
spectrum $\mathcal{P}_{\mathcal{R}}^{\mathrm{QT}}(k)$ is scale invariant. From
(\ref{Cl2}) and (\ref{xi2}) we then have an approximate ratio
\begin{equation}
\frac{C_{l}}{C_{l}^{\mathrm{QT}}}=2l(l+1)\int_{0}^{\infty}\frac{dk}{k}
\ j_{l}^{2}(2k/H_{0})\xi(k) \label{ratio}
\end{equation}
(where $C_{l}^{\mathrm{QT}}$ is the quantum-theoretical angular power
spectrum).\footnote{We ignore the small contribution from the integrated
Sachs-Wolfe effect at very low $l$.} A low power anomaly $C_{l}/C_{l}
^{\mathrm{QT}}<1$ in the CMB may be explained by an appropriate primordial
deficit $\xi(k)<1$ \cite{AV10,CV13}.

If the primordial deficit takes the inverse-tangent form (\ref{fit1}), we may
evaluate the expression (\ref{ratio}) numerically to find the range of
parameters $c_{1}$, $c_{2}$, $c_{3}$ giving an angular power deficit in the
(crudely speaking) observed range $\sim5-10\%$ -- that is, giving a ratio
$C_{l}/C_{l}^{\mathrm{QT}}$ in the range $0.9-0.95$. (Of course the deficit
reported by the Planck team is a statistical aggregate for the whole low-$l$
region and does not refer to individual multipoles, so this is only a rough
characterisation of the data.)

It is convenient to use the variable $x\equiv2k/H_{0}$ and to define a
rescaled coefficient
\begin{equation}
\tilde{c}_{1}\equiv(H_{0}/2\pi)c_{1}\ .
\end{equation}
We then have
\begin{equation}
\frac{C_{l}}{C_{l}^{\mathrm{QT}}}=2l(l+1)\int_{0}^{\infty}\frac{dx}{x}
\ j_{l}^{2}(x)\left(  \tan^{-1}(\tilde{c}_{1}x+c_{2})-\frac{\pi}{2}
+c_{3}\right)  \label{ratio2}
\end{equation}
(where $j_{l}^{2}(x)$ is dominated by the scale $x\simeq l$). We have
evaluated this expression numerically at low $l$ for varying values of
$\tilde{c}_{1}$, $c_{2}$ but keeping $c_{3}$ fixed at $c_{3}=1$. Our results
are trivially extended to arbitrary $c_{3}$, since writing $c_{3}=1+(c_{3}-1)$
the expression (\ref{ratio2}) takes the form
\[
C_{l}/C_{l}^{\mathrm{QT}}=\left(  C_{l}/C_{l}^{\mathrm{QT}}\right)  _{c_{3}
=1}+(c_{3}-1)
\]
where we have used $\int_{0}^{\infty}\frac{dx}{x}\ j_{l}^{2}(x)=1/2l(l+1)$.

\begin{figure}
\begin{center}
\includegraphics[width=0.9\textwidth]{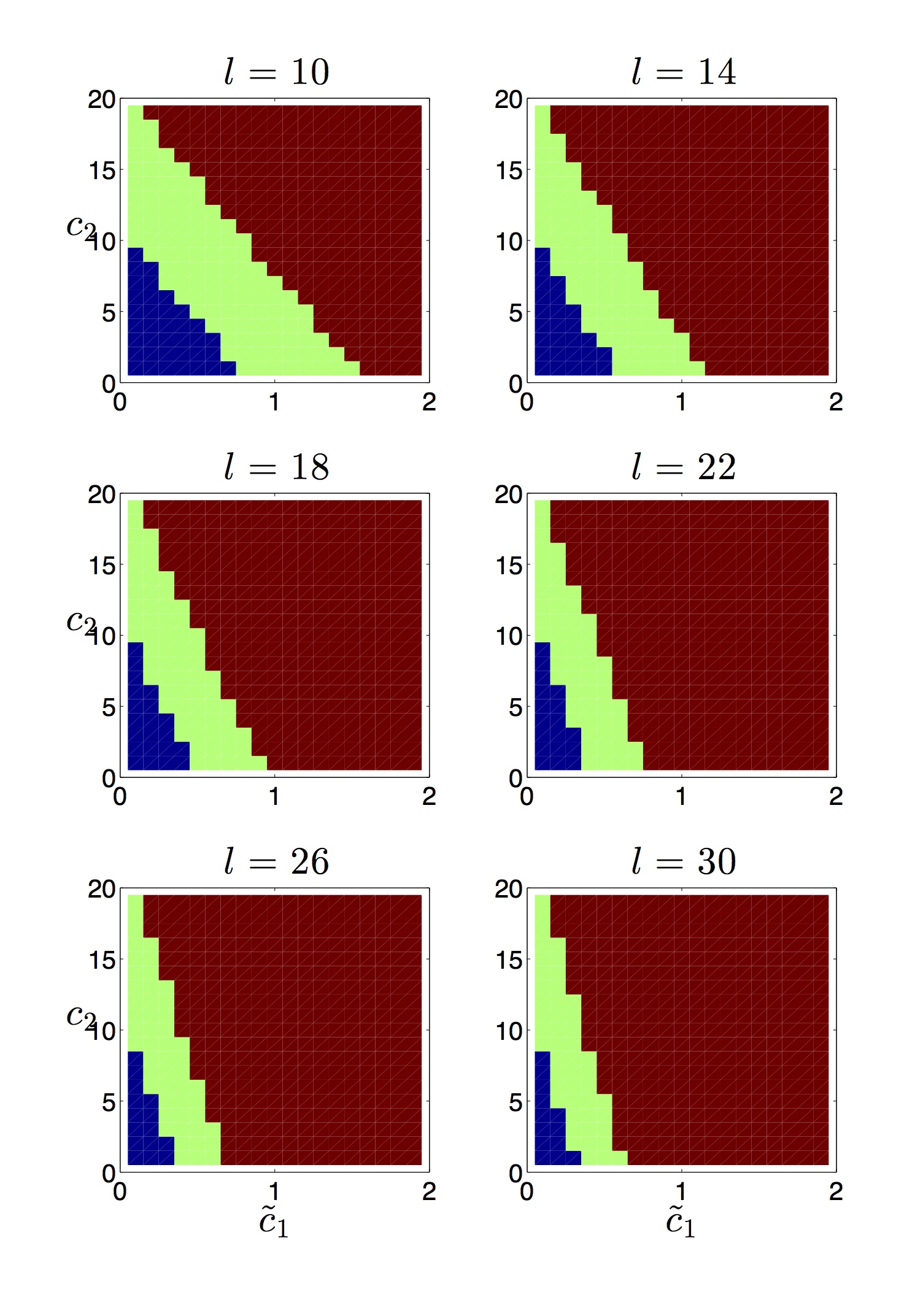}
\caption{\label{fig16}Parameter space $(\tilde{c}_{1},c_{2})$ for $l=10,14,18,22,26,30$.
The green regions have $C_{l}/C_{l}^{\mathrm{QT}}$ in the range $0.9-0.95$;
the blue regions have $C_{l}/C_{l}^{\mathrm{QT}}<0.9$; the red regions have
$C_{l}/C_{l}^{\mathrm{QT}}>0.95$. (Each plot uses $20\times20$ grid points.)}
\end{center}
\end{figure}

Plots of the calculated deficit $C_{l}/C_{l}^{\mathrm{QT}}$ on the parameter
space $(\tilde{c}_{1},c_{2})$ are displayed in Figure \ref{fig16} for
$l=10,14,18,22,26,30$. The green region corresponds to $C_{l}/C_{l}
^{\mathrm{QT}}$ in the range $0.9-0.95$. (The blue region corresponds to
$C_{l}/C_{l}^{\mathrm{QT}}<0.9$ while the red region corresponds to
$C_{l}/C_{l}^{\mathrm{QT}}>0.95$.) The green `deficit region' has parameters
$\tilde{c}_{1}$ in the approximate range $[0,1.5]$ and $c_{2}$ in the
approximate range $[0,20]$ (restricting ourselves to $\tilde{c}_{1},c_{2}
>0$).\footnote{Note that we have mapped only a part of the possible parameter
space for $(\tilde{c}_{1},c_{2})$. This suffices for our present purposes.}

Of course we have not performed a best-fit, we have simply obtained the
magnitudes that $\tilde{c}_{1}$ and $c_{2}$ must have in order for the low-$l$
angular spectrum to drop by $5-10\%$.

\section{Comparing with observation}

Let us now compare the required ranges for $\tilde{c}_{1},c_{2}$ -- as deduced
from the required angular power deficits -- with the results for $c_{1},c_{2}$
obtained from our model. (We will not consider $c_{3}$ here as it may be
reabsorbed into an overall renormalisation of the primordial power spectrum.)

To relate our results to observable quantities, we must take into account the
spatial expansion by a factor
\begin{equation}
R\equiv\frac{a(t_{\mathrm{today}})}{a(t_{f})}
\end{equation}
that will have taken place from the end of pre-inflation at time $t_{f}$ until
today at time $t_{\mathrm{today}}$. The true coefficient $c_{1}$ (or
$\tilde{c}_{1}$) that appears in the observable power spectrum is then
multiplied by the unknown number $R$.

Let us consider this last point more carefully. In our simulations we used
natural units ($\hbar=c=1$) with fiducial initial and final times
$t_{i}=10^{-4}$, $t_{f}=10^{-2}$. For convenience we also took $a_{0}=1$ at
$t_{0}=1$. This means that in our simulations -- which yielded a deficit
function $\xi(k)$ of the form (\ref{fit1}) -- $k$ was the physical wavenumber
at time $t=t_{0}=1$ (in our units). During the simulated pre-inflationary era
we have $a(t)=a_{0}(t/t_{0})^{1/2}=t^{1/2}$ and so (as already noted) at the
end of pre-inflation our scale factor is $a_{f}=t_{f}^{1/2}=0.1$. Thus
physical wavenumbers at the final time $t_{f}$ are ten times larger than the
physical wavenumbers $k$ referred to in our simulations. It follows that
physical wavenumbers today, which we might temporarily denote as
$k_{\mathrm{today}}$, are given by
\begin{equation}
k_{\mathrm{today}}=(10/R)k\ .
\end{equation}
Now, if we assume that our simulated deficit function $\xi(k)$ survives the
transition to inflation and indeed provides a correction to the inflationary
power spectrum, then the same deficit will enter as a factor in the spectrum
of primordial perturbations. It follows that the true correction
$\xi_{\mathrm{today}}(k_{\mathrm{today}})$ multiplying the observable
primordial spectrum will be
\begin{equation}
\xi_{\mathrm{today}}(k_{\mathrm{today}})=\left[  \xi(k)\right]
_{k=(R/10)k_{\mathrm{today}}}
\end{equation}
(since the mode which we today label $k_{\mathrm{today}}$ was the mode
labelled $k=(R/10)k_{\mathrm{today}}$ in our simulations). We may then drop
the subscript `today' and simply write the true deficit correction as $\xi(k)$
with $k$ now denoting physical wavenumbers today (as is more conventional). In
other words, the true correction multiplying the observable primordial
spectrum will be numerically equal to our simulated function $\xi(k)$ but with
$k$ replaced by $(R/10)k$ and with $k$ then reinterpreted as the physical
wavenumber today. With this understanding, we obtain a true deficit function
\begin{equation}
\xi(k)=\tan^{-1}(\tilde{c}_{1}x+c_{2})-\frac{\pi}{2}+c_{3} \label{trueksi}
\end{equation}
with $x=2k/H_{0}$ and
\begin{equation}
\tilde{c}_{1}\equiv c_{1}RH_{0}/20\pi\ . \label{truecoeff}
\end{equation}

Thus the deficit in the angular power spectrum will still take the form
(\ref{ratio2}) but with an observed coefficient $\tilde{c}_{1}$ given by
(\ref{truecoeff}), where $c_{1}$ is the coefficient generated by our numerical
simulations of the pre-inflationary era (using our convenient units).

We have seen that our simulations yield (to a first approximation)
coefficients $c_{1}$ and $c_{2}$ that depend on both $M$ and $t_{f}$ and a
coefficient $c_{3}$ that depends only on $M$. The observable spectra will be
sensitive to $c_{1}$ via the factor $c_{1}R$ appearing in (\ref{truecoeff}),
and so uncertainty in $M$ and $t_{f}$ (and hence in $c_{1}$) is compounded
with uncertainty in $R$. On the other hand, the spectra are (at least in
principle)\ directly sensitive to $c_{2}$, $c_{3}$ -- which are determined by
the dynamics of our model (for a given number $M$ of pre-inflationary excited
states and for a given duration $(t_{i},t_{f})$, and assuming the simple
initial nonequilibrium distribution (\ref{rhoinitial})). The spectra are also
in principle directly sensitive to the inverse-tangent functional form of
$\xi(k)$, a feature that is again determined by the dynamics of our model and
which seems to be a robust prediction for a fairly broad range of parameters
$(t_{f},M)$ characterising the pre-inflationary era. Our simulations also
predict phase anomalies at comparable values of $k$, though the implications of
this for observable quantities (such as signatures of anisotropy) remain to be explored.

We are unable to predict $R$, but it is subject to well-known constraints.
Should $R$ be too large, the effective coefficient $\tilde{c}_{1}$ as given by
(\ref{truecoeff}) could be so large that our deficit function (\ref{trueksi})
appearing in (\ref{ratio2}) will be essentially equal to the constant $c_{3}$
and will therefore simply generate an overall renormalisation of the power
spectrum. If more interesting features of $\xi(k)$ are to be observable, we
must assume that the value of $\tilde{c}_{1}$ is such that the $k$-dependence of 
$\xi(k)$ occurs in an observable range of $k$-space. If we are fortunate and
this is correct, we will then be able to test the other (predicted) details of
the function $\xi(k)$ -- its functional dependence on $k$, and the values of
the other two coefficients $c_{2}$, $c_{3}$.

The first thing to check is whether or not our model can yield the required
angular power deficit -- with $C_{l}/C_{l}^{\mathrm{QT}}$ within the `green
zone' of Figure \ref{fig16} -- for reasonable values of the expansion factor $R$. It
suffices to find an example of our predicted coefficients $c_{1}$, $c_{2}$
(for some pair of parameters $t_{f}$, $M$) that corresponds to the observed
deficit region for an acceptable choice of cosmological parameters.

Our simulations with fixed $t_{f}=10^{-2}$ and varying $M$ (Section 3.1)
yielded $c_{1}$ in the range $[0.10,0.46]$ and $c_{2}$ in the range
$[0.83,2.85]$. Let us consider the `preferred' case with the smallest number
$M=4$ of excited states, which yielded $c_{1}=0.46$ and $c_{2}=2.85$ (see
Table \ref{table1}). If we could take $\tilde{c}_{1}\simeq0.5$ then from Figure \ref{fig16} we
see that the point $(\tilde{c}_{1},c_{2})\simeq(0.5,2.85)$ would lie in or
close to the green zone for all the displayed multipoles $l$. To obtain
$\tilde{c}_{1}\simeq0.5$ when $c_{1}=0.46$ requires (from (\ref{truecoeff}))
$RH_{0}/20\pi\sim1$. Now our coefficient $c_{1}$ has conventional units of
length, or mass dimension $-1$ in natural units, and so our numerical values
$c_{1}=x$ must be multiplied by $\hbar c$ to obtain values $c_{1}
\simeq(3\times10^{-24}\ \mathrm{cm})x$ in conventional units. Thus our
requirement $RH_{0}/20\pi\sim1$ is really $(3\times10^{-24}\ \mathrm{cm}
)RH_{0}/20\pi\sim1$ or
\begin{equation}
R\sim10^{53} \label{con}
\end{equation}
(where $H_{0}^{-1}\simeq10^{28}\ \mathrm{cm}$).

Is (\ref{con}) consistent with known cosmological constraints? To see that it
is, let us write $R$ as
\[
R\equiv a_{\mathrm{today}}/a_{f}=(a_{\mathrm{today}}/a_{\mathrm{end}
})(a_{\mathrm{end}}/a_{f})\ ,
\]
where $a_{\mathrm{end}}$ is the scale factor at the end of inflation. We may
neglect the expansion that takes place during the transition (from
pre-inflation to inflation) compared to the expansion that takes place during
inflation. Thus we may approximately identify $a_{f}$ with the scale factor
$a_{\mathrm{begin}}$ at the beginning of inflation. We then have
$a_{\mathrm{end}}/a_{f}\simeq e^{N}$ where $N$ is the number of inflationary
e-folds. Let us similarly neglect the expansion that takes place during the
transition from inflation to post-inflation. We can then write
$a_{\mathrm{today}}/a_{\mathrm{end}}\simeq T_{\mathrm{end}}/T_{\mathrm{today}
}$ (where $T_{\mathrm{end}}$ is the temperature at which inflation ends). We
then find%
\begin{equation}
R\simeq e^{N}(T_{\mathrm{end}}/T_{\mathrm{today}})\ . \label{R}
\end{equation}
The value of the reheating temperature $T_{\mathrm{end}}$ depends on the
details of the reheating process. (See for example refs. \cite{PU09,Alla10}.)
We certainly have an upper bound $T_{\mathrm{end}}\lesssim T_{f}$, where
$T_{f}$ is the temperature at the end of pre-inflation. Indeed we could even
have $T_{\mathrm{end}}<<T_{f}$. We may expect $T_{f}$ to be comparable to the
energy scale $H_{\mathrm{\inf}}\sim10^{16}\ \mathrm{GeV}$ of inflation (where
$H_{\mathrm{\inf}}$ is the Hubble parameter during inflation). Thus we may
safely write an upper bound $T_{\mathrm{end}}\lesssim10^{16}\ \mathrm{GeV}$.
Lower bounds on $T_{\mathrm{end}}$ in the range $390\ \mathrm{GeV}
-890\ \mathrm{TeV}$ (depending on the inflationary model) have been obtained
from CMB data \cite{MR10}. Thus we may take a rough lower bound
$T_{\mathrm{end}}\gtrsim1\ \mathrm{TeV}$. We then have $10^{16}
<T_{\mathrm{end}}/T_{\mathrm{today}}<10^{29}$ (where $T_{\mathrm{today}}
\sim10^{-4}\ \mathrm{eV}$), which from (\ref{R}) implies the bounds
$10^{16}e^{N}<R<10^{29}e^{N}$. Our condition (\ref{con}) then implies a range
\begin{equation}
55<N<85\ .
\end{equation}
This is compatible with standard constraints, which indicate that the minimum
number $N=N_{\min}$ of e-folds (required for inflation to solve the horizon
and flatness problems) is $N_{\min}\simeq70$ -- with some authors taking
$N_{\min}\simeq60$. See, for example, ref. \cite{PU09}. (The actual number $N$
of e-folds could of course be much larger than $N_{\min}$. As we have noted,
if $N$ is too large then the power deficit generated by our model would exist
at wavelengths too large to be observable.) Thus there exist acceptable values
for the cosmological parameters $N$, $T_{\mathrm{end}}$ such that (\ref{con})
is satisfied, in which case the simulated coefficients $c_{1}=0.46$,
$c_{2}=2.85$ will be consistent with the observed deficit (corresponding to
the green zone of Figure \ref{fig16}).

We conclude that the range of values for $c_{1}$, $c_{2}$ required by
observation is compatible with the range of values for $c_{1}$, $c_{2}$
obtained from our relaxation simulations. Thus we may say that our model seems
viable -- pending a full treatment of the transition to inflation.

\section{Conclusion}

Primordial quantum relaxation provides a single mechanism that can generate
both a power deficit and anomalous phases at large angular scales in the CMB.
Our estimates show that, with an appropriate choice of cosmological
parameters, our model is able to generate a power deficit at approximately the
angular scales and of approximately the magnitude that has been reported by
the Planck team, as well as generating anomalous phases at comparable angular
scales. In addition, the same mechanism generates oscillations in the
primordial spectrum.

There are of course other mechanisms that can produce a large-scale power
deficit, such as a suitable period of inflationary `fast rolling' \cite{CL03}.
It is hoped that the particular form of our power deficit $\xi(k)$, given by
(\ref{trueksi}) as a function of wavenumber $k$, will distinguish it from the
deficit predicted by alternative models.

From the viewpoint of our underlying model, it must be assumed that the number
of inflationary e-folds is not too large, for otherwise our effects would
exist at wavelengths that are too large to be detectable. On the other hand,
once this assumption is made our model makes several clear and testable
predictions: an inverse-tangent correction $\xi(k)$ to the large-scale
primordial spectrum, with oscillations around the curve, and with anomalous
phases at comparable scales. Of the three parameters $\tilde{c}_{1}$, $c_{2}$,
$c_{3}$ appearing in our fit (\ref{trueksi}), the first depends on the number
$N$ of inflationary e-folds and on the inflationary reheating temperature
$T_{\mathrm{end}}$; but the second and third are entirely determined by our
model. The parameters $\tilde{c}_{1}$, $c_{2}$, $c_{3}$ depend on the final
time $t_{f}$ and the number $M$ of excited energy states (both defined for the
pre-inflationary era). An alternative fit (\ref{oscfit1}), involving an
exponential of the retarded time $t_{\mathrm{ret}}(t_{f},k)$, provides an
equally good fit to the overall shape of the curve while also capturing some
features of the oscillations in the low-$k$ region.

In effect, then, we have a model of the power deficit in terms of two
parameters $t_{f}$, $M$ -- for a given initial time $t_{i}$ and a given
initial nonequilibrium distribution (\ref{rhoinitial}). The model generates
some rather complex features: a power deficit of a particular form, with
oscillations around the curve and with anomalous phases. The prospects are
therefore good for a comparison with data. It is of course quite possible that
the number of inflationary e-folds is so large that, even if our effects
exist, they will be too faint to be observable. But if the effects are
visible, they should show detailed signatures.

In this paper we have assumed a fixed initial nonequilibrium distribution
(\ref{rhoinitial}), equal to the equilibrium distribution for the ground state
of the field mode. This was chosen as a simple example of a nonequilibrium
distribution whose initial width is smaller than the initial quantum width.
One may ask to what extent our final results depend on this choice. We may
reasonably expect to find similar results if the initial nonequilibrium
distribution is a simple smooth function whose width is smaller than the
quantum width (for example, a function equal to the equilibrium distribution
associated with a superposition of $M^{\prime}<M$ energy states). For the
erratic motion of the trajectories is likely to erase any dependence on the
finer details of the initial distribution. We are after all only concerned
with the final width -- other details of the final distribution do not affect
our calculation of the power deficit. We may therefore expect that our results
will depend mainly on the initial width only, and not on other details of the
initial distribution. This expectation is confirmed by further simulations, in
which the initial nonequilibrium distribution includes terms that in quantum
equilibrium would result from the first excited state of the field mode. We
have found that, while the details of the final distribution are slightly
different, the final power deficit has the same inverse-tangent dependence on
wavelength as before (with slightly different best-fit parameters). These
further simulations will be reported elsewhere, in a separate publication in
which we also study the effect of the transition \cite{CV15}.

The extent to which the data support our model remains to be seen. A first
step would be to evaluate likelihoods for corrections to the power spectrum of
the form (\ref{trueksi}) \cite{PVV15}. The class of models to be fitted will
include the standard cosmological parameters together with our extra
parameters. A second step would be to evaluate the extent to which our
predicted oscillations are present in the data. Finally, one may also consider
in more detail the apparent large-scale anisotropy in the CMB and how it might
be explained by the anomalous primordial phases that our model suggests could
exist at very large scales.

\textbf{Acknowledgements}. AV wishes to thank Patrick Peter for helpful
discussions. We are grateful to Murray Daw for kindly providing us with extra
computational resources on the Clemson University Palmetto Cluster. This
research was funded jointly by the John Templeton Foundation and Clemson University.

\end{document}